\documentclass[prd,twocolumn,superscriptaddress,floatfix,nofootinbib]{revtex4-2}
\pdfoutput=1 % if your are submitting a pdflatex (i.e. if you have
\usepackage[T1]{fontenc}
\usepackage{amsmath}
\usepackage{amssymb}
\usepackage{amsthm}
\usepackage{amsfonts}
\usepackage{graphicx}
\usepackage{subcaption}
\usepackage{enumitem}
\usepackage{bm}
\usepackage{rotating}
\usepackage[colorlinks]{hyperref}
 \hypersetup{
     colorlinks=true,
     linkcolor=BlueViolet,
     filecolor=blue,
     citecolor=BrickRed, % Hyperref citation colour
     urlcolor=cyan,
     }
\usepackage{pbox}
\usepackage[dvipsnames,usenames]{xcolor}
\usepackage{array}
\usepackage{physics}
\usepackage{dsfont}
\usepackage{mathtools}
\usepackage{leftidx}
\usepackage{lmodern}
\usepackage[normalem]{ulem}
\usepackage{braket}
\usepackage{tensor}
\usepackage{mathrsfs}
\usepackage{float}
\usepackage{dsfont}
\usepackage[margin=2cm]{geometry}
\usepackage[title]{appendix}
\usepackage{tcolorbox}
\usepackage{tocloft}
\usepackage{titlesec}
\usepackage{ragged2e}%used for centering caption
\usepackage{tikz}
\usetikzlibrary{arrows.meta,decorations.pathmorphing,math}

\definecolor{darkraspberry}{rgb}{0.53, 0.15, 0.34}
\definecolor{mypink}{RGB}{226,68,130}

\newcommand{\ii}{\mathsf{i}}

\newcommand{\sx}{\boldsymbol{\mathsf{x}}}

\newcommand{\D}{\text{d}}

\begin{document}

\title{Accelerated detector in a superposed spacetime}

\author{Lakshay Goel}
\email{l2goel@uwaterloo.ca}
\affiliation{Department of Physics and Astronomy, University of Waterloo, Waterloo, Ontario, N2L 3G1, Canada}

\author{Everett A. Patterson}
\email{ea2patterson@uwaterloo.ca}
\affiliation{Department of Physics and Astronomy, University of Waterloo, Waterloo, Ontario, N2L 3G1, Canada}
\affiliation{Institute for Quantum Computing, University of Waterloo, Waterloo, Ontario, N2L 3G1, Canada}
\affiliation{Waterloo Centre for Astrophysics, University of Waterloo, Waterloo, ON, N2L 3G1, Canada}
\affiliation{Perimeter Institute for Theoretical Physics, 31 Caroline St N, Waterloo, Ontario, N2L 2Y5, Canada}

\author{Mar\'ia Rosa Preciado-Rivas}
\email{mrpreciadorivas@uwaterloo.ca}
\affiliation{Department of Physics and Astronomy, University of Waterloo, Waterloo, Ontario, N2L 3G1, Canada}
\affiliation{Institute for Quantum Computing, University of Waterloo, Waterloo, Ontario, N2L 3G1, Canada}
\affiliation{Waterloo Centre for Astrophysics, University of Waterloo, Waterloo, ON, N2L 3G1, Canada}

\author{Mahdi Torabian}
\email{mahdi.torabian@gmail.com}
\affiliation{Perimeter Institute for Theoretical Physics, 31 Caroline St N, Waterloo, Ontario, N2L 2Y5, Canada}

\author{Robert B. Mann}
\email{rbmann@uwaterloo.ca}
\affiliation{Department of Physics and Astronomy, University of Waterloo, Waterloo, Ontario, N2L 3G1, Canada}
\affiliation{Institute for Quantum Computing, University of Waterloo, Waterloo, Ontario, N2L 3G1, Canada}
\affiliation{Waterloo Centre for Astrophysics, University of Waterloo, Waterloo, ON, N2L 3G1, Canada}
\affiliation{Perimeter Institute for Theoretical Physics, 31 Caroline St N, Waterloo, Ontario, N2L 2Y5, Canada}

\author{Niayesh Afshordi}
\email{nafshordi@uwaterloo.ca}
\affiliation{Department of Physics and Astronomy, University of Waterloo, Waterloo, Ontario, N2L 3G1, Canada}
\affiliation{Waterloo Centre for Astrophysics, University of Waterloo, Waterloo, ON, N2L 3G1, Canada}
\affiliation{Perimeter Institute for Theoretical Physics, 31 Caroline St N, Waterloo, Ontario, N2L 2Y5, Canada}

\date{\today}

\begin{abstract}

In pursuit of a full-fledged theory of quantum gravity, operational approaches offer insights into quantum-gravitational effects produced by quantum superposition of different spacetimes not diffeomorphic to one another. Recent work applies this approach to superpose cylindrically identified Minkowski spacetimes (i.e. periodic boundary conditions) with different characteristic circumferences, where a two-level detector coupled to a quantum field residing in the spacetime exhibits resonance peaks in response at certain values of the superposed length ratios. Here, we extend this analysis to a superposition of cylindrically identified Rindler spacetimes, considering a two-level detector constantly accelerated in the direction orthogonal to the compact dimension. Similarly to previous work, we find resonance peaks in the detector response at rational ratios of the superposed compactified lengths, which we observe to be accentuated by the acceleration of the detector. Furthermore,  for the first time
we confirm the detailed balance condition due to  acceleration  in a superposition of spacetimes,
commensurate with the  Unruh effect in a single spacetime state.
The resonant structure of detector response in the presence of event horizons, for the first time observed in 3+1 dimensions, may offer clues to the nature of black hole entropy in the full theory of quantum gravity. 
\end{abstract}

 \maketitle

\section{Introduction}

Despite lacking a complete theory,  there has been interest in studying the phenomenology of quantum gravity from an operational viewpoint~\cite{Rideout:2012jb,Altamirano:2016fas,Asenbaum:2024esg,Overstreet:2021hea,Overstreet:2022zgq,Overstreet:2017gdp,danielson2022black,arrasmith2019,Demers_1996,allaliPhysRevLett.127.031301,Kovachy:2015xcp,bose2017spin,marletto2017gravitationally}.
This approach employs (theoretical) physical devices such as detectors, rods, and clocks that ground observables with in-principle measurements.
Recent advancements in the field of relativistic quantum information \cite{Ralph:2012mdp} and quantum field theory in curved spacetimes have enabled us to combine notions of entanglement, superposition, and measurement from fundamentals of quantum theory with notions from general relativity, such as proper time, causal structure, and spacetime.
In contrast to the more abstract methods used in the `top-down' approaches that restructure theoretical foundations and redefine quantum spacetime, the `bottom-up' operational approach focuses on empirically accessible phenomena that are expected to generically arise from quantum-gravitational physics.

In this paper we study a generic situation expected from any theory of quantum gravity, namely the quantum superposition of spacetimes. We work within a recently developed framework~\cite{FMZ_2021_dS, Schordinger_BH_cat,FAZM_2022_BTZ, FAZM_2023_flat, Diki_2023_BTZ_rotate}
based on quantum reference frames~\cite{2004SHPMP..35..195D,Giacomini_2019,Giacomini_2022}.
Explicit quantum superposition of spacetimes can be understood to arise as follows~\cite{FMZ_2021_dS}:   a measuring device placed in a superposition of two different spacetime trajectories (for example, due to a Mach-Zehnder interferometer)  can be mapped to a description in which the device is placed in a superposition of two spacetimes that are diffeomorphic to one another. This map can be inverted, so the two descriptions are equivalent.  However, this map can also be extended to situations in which the two spacetimes are not related by a diffeomorphism and are instead distinct solutions to Einstein's field equations. In this case, the quantum state of a single device placed in a superposition of two such spacetimes is no longer equivalent to the quantum state of a device in a superposition of trajectories in a single classical background spacetime.  
Such a description of spacetime superposition does not constitute nor require a full quantum-gravitational theory; rather, it only employs the assumption that such superpositions belong to the Hilbert space of allowed states of such a theory.

Recently the superposition of two non-rotating chargeless Bañados-Teitelboim-Zanelli (BTZ) black holes \cite{BTZ_BH} of different masses was
constructed, and the response of a detector in such a setting was subsequently computed \cite{FAZM_2022_BTZ}. Specifically, the response of a quantum detector interacting with a massless quantum scalar field that is conformally coupled to the superposed spacetime was found to   peak at the rational ratios of the horizon radii in each branch of the superposition, a result that is consistent with Bekenstein's area quantization conjecture \cite{Bekenstein_1, Bekenstein_2,Bekenstein:1995ju}. Similar results were subsequently obtained for rotating BTZ black holes \cite{Diki_2023_BTZ_rotate}.

While modeling BTZ black holes gives us insight in 2+1 dimensions, we are ultimately interested in understanding quantum gravitational effects in (3+1)-dimensional spacetime and how they might be detected. For example, horizon area quantization may lead to gravitational wave echoes, potentially observable in the aftermath of binary black hole mergers (e.g. \cite{Cardoso:2019apo,Agullo:2020hxe}). However, working with a superposition of (3+1)-dimensional black holes presents significant computational challenges for
computing detector response
\cite{Ng:2021enc}.
Fortunately, we can gain some insight by considering simpler situations. A recent example considers a static quantum detector interacting with a massless scalar quantum field minimally coupled to two superposed cylindrically identified (3+1)-dimensional Minkowski spacetimes of different characteristic circumferences \cite{FAZM_2023_flat}. The response of a static detector in this setting was found to be sensitive to the spacetime superposition, with resonances occurring at rational values of the ratio of the circumferences of the superposed cylinders.

Left unanswered is   the question of how these results are affected by  detector motion.
In particular, accelerated detectors provide key insights connecting causal structures to quantum field theoretic effects. The best known is the Unruh effect, which predicts that a uniformly accelerated detector in the Minkowski vacuum,
exhibits a thermal response \cite{Unruh_effect_1976}, with a temperature $T_U = \frac{a}{2\pi}$  where $a$ is the acceleration of the detector.

In the \textit{Unruh-DeWitt} (UDW) detector \cite{udw_detector} framework, where a 2-state point-like detector couples to a quantum field,
the Unruh effect is characterized by the \textit{excitation-to-deexcitation ratio} (EDR) satisfying the \textit{detailed balance condition} (DBC) in the late-time equilibrium limit \cite{Lima_Everett_2023, Carballo_Rubio_Edu_2019}, i.e., 
\begin{equation}
    \lim_{t \to \infty} \frac{\text{P}(\Omega)}{\text{P}(-\Omega)} = e^{-\beta\Omega} = e^{-2\pi\Omega/a},
    \label{eq:detailed balance condition}
\end{equation}
where $\text{P}(\Omega)$ ($\text{P}(-\Omega)$) represents the detector excitation (deexcitation) probability, $\Omega$ is the energy gap between detector's internal states, and $\beta = T_U^{-1} = \frac{2\pi}{a}$ is the inverse temperature.

More recent investigations have shown that an
\textit{Unruh-DeWitt} (UDW) detector \cite{udw_detector} placed in a superposition of uniformly accelerated trajectories (with differing Rindler horizons) does not register a thermal response \cite{FOZ_2020_suptrajUDW}.
Here, we study an analogous problem inspired by expectations of quantum gravity: what is the response of a uniformly accelerated UDW detector when placed in a superposed spacetime?

To address this question, we consider a superposition of two (3+1)-dimensional flat spacetimes with different cylindrical identifications and
compute the response of a uniformly accelerated quantum detector, linearly coupled to a massless scalar field,  in this setting.
Similar to the BTZ case \cite{FAZM_2022_BTZ}, we find that the response exhibits resonance peaks that depend on the ratio of the characteristic circumferences of the cylinders. We also investigate the validity of the \textit{detailed balanced condition} (DBC) for the Rindler temperature of the superposed spacetime.

Our paper is organized as follows. In Sec. \ref{sec:spacetime-field}, we review the Rindler spacetime and its cylindrical topological identification. We then provide a description of a massless scalar field in this spacetime, along with its two-point correlation function.
In Sec. \ref{sec:detector} we introduce the quantum controlled UDW model, construct the density matrix for a uniformly accelerated detector in the superposed spacetime setup, and compute
the formulae for the transition probability of the UDW detector.
We note that thermality can be described via the DBC for these transition probabilities \cite{Carballo_Rubio_Edu_2019, Lima_Everett_2023}.
In Sec. \ref{sec:results}, we present our results for the detector response as a function of the energy gap of the detector and the ratio of the characteristic circumferences of the two branches of the superposition.  We conclude with some final thoughts in Sec. \ref{sec:conclusion}. Throughout this paper, we use natural units $\hbar = k_B = c = 1$.

\section{Spacetime and Fields}\label{sec:spacetime-field}

In this section, we review the basic structure of the Rindler spacetime and the Minkowski vacuum, before describing how they are affected by periodic identifications in the $z$ coordinate.

\subsection{Rindler coordinates}
\label{sec:rindler}

The line element for (3+1)-dimensional Minkowski spacetime, $\mathcal{M}$,
parameterized by coordinates $\sx \equiv (t, x, y, z)$ is given by
\begin{equation}
    \D s^2 = \D t^2 - \D x^2 - \D y^2 - \D z^2,
    \label{eq:line-element-minkowski}
\end{equation}
with  metric signature $(+, -, -, -)$  \cite{FAZM_2023_flat, EduAlexDani2016}.
Transforming to \textit{Rindler coordinates}, $t = r \sinh{(a\tau)}$, $x = r \cosh{(a\tau)}$, $y=y$, $z=z$, where $a$ is the proper acceleration and $\tau$ is the proper time of a uniformly accelerated observer, we express the line element for the right Rindler wedge, $\mathcal{R}$, as
\begin{equation}
    \D s^2 = r^2a^2\D \tau^2 - \D r^2 - \D y^2 - \D z^2, 
    \label{eq:line-element-rindler}
\end{equation}
where a uniformly accelerated trajectory corresponds to
setting $r = \frac{1}{a}$. 

The geodesic distance squared between two spacetime points in Minkowski coordinates is
\begin{equation}
    \sigma (\sx, \sx ') = (t-t')^2 - (x-x')^2 - (y-y')^2 -(z-z')^2, 
    \label{eq:geodesic-distance-minkowski}
\end{equation}
which is
\begin{equation}
    \sigma_\mathcal{R} (\sx, \sx ') = \frac{4}{a^2} \sinh^2{\left[\frac{a}{2}(\tau - \tau ')\right]} -(y-y')^2 -(z-z')^2,
    \label{eq:geodesic-distance-rindler}
\end{equation}
in Rindler coordinates, as shown in \autoref{appendix:p_d,l_ab}.

We consider a massless scalar field $\hat{\phi}$ that is a solution to the Klein-Gordon equation $\Box\, \hat{\phi}(\sx) = 0$, where $\Box$ is the d'Alembert operator. The field operator $\hat{\phi}$ is then given by
\begin{equation}
    \hat{\phi} (\sx) = \int \frac{\D^3 k}{(2\pi)^{3/2}} \frac{1}{\sqrt{2|\textbf{k}|}} \left(e^{-i |\textbf{k}| t + i \textbf{k} \cdot\textbf{x} } \hat{a}_\textbf{k} + \text{H.c.}\right),
    \label{eq:klein-gordon-phi}
\end{equation}
where $\textbf{k} = (k_x, k_y, k_z)$, $\textbf{x} = (x,y,z)$ are the momentum and position 3-vectors respectively,  $\hat{a}_\textbf{k} (\hat{a}_\textbf{k}^{\dag})$ is the annihilation (creation) operator for a single-frequency mode, and H.c. denotes the Hermitian conjugate.

The two-point correlation function pulled back to the worldlines $(\sx, \sx ')$, commonly referred to as the Wightman function, is given by $W(\sx, \sx') := \langle 0| \hat{\phi}(\sx)\hat{\phi}(\sx ')|0\rangle $, where $\ket{0}$ is the Minkowski vacuum state which is annihilated by $\hat{a}_\textbf{k}$~\cite{birrell_davies_1982}. 
Expressed in Rindler coordinates, this is   \cite{FAZM_2023_flat, EduAlexDani2016}
\begin{equation}
    \begin{split}
        W_{\mathcal{R}}(\sx, \sx ') = \frac{1}{4\pi i} \text{sgn}(\tau-\tau') \delta(\sigma_\mathcal{R}(\sx, \sx ')) - \frac{1}{4 \pi^2 \sigma_\mathcal{R}(\sx, \sx ')},
    \end{split}
    \label{eq:W_R}
\end{equation}
where sgn$(\tau-\tau') = \pm 1$ depends on the sign of $\tau-\tau'$, and $\sigma_\mathcal{R}(\sx, \sx')$ is the geodesic distance squared given by Eq. (\ref{eq:geodesic-distance-rindler}).

\subsection{The $\mathcal{R}/J_0$ Quotient Space}
\label{sec:quotient-space}

Let us now consider a quotient of Minkowski space in Rindler coordinates with non-trivial topology. The cylindrical universe $R_0 = \mathcal{R}/J_0$ is constructed as a quotient of the Rindler wedge $\mathcal{R}$ under the isometry group $Z \simeq J_0^n, \,J_0:(\tau,r,y,z) \mapsto (\tau,r,y,z+l)$ with characteristic circumference $l$ \cite{FAZM_2023_flat}. $J_0$ acts freely, thus ensuring that $R_0$ is a space and time orientable Lorentzian manifold. 
This isometry generates a cylindrically identified Minkowski spacetime (expressed in Rindler coordinates).

To construct a quantum field theory on this background, consider an automorphic field $\hat{\psi}(\sx)$ constructed from the massless scalar field $\hat{\phi}(\sx)$ as the image sum \cite{Langlois_2006}
\begin{equation}
    \hat{\psi}(\sx) := \frac{1}{\sqrt{\mathcal{N}}}\sum_{n=-\infty}^{\infty} \eta^n \hat{\phi}(J_0^n \sx),
    \label{eq:qft automorphic field}
\end{equation}
where $\mathcal{N} = \sum_n \eta^{2n}$ is the normalization factor that ensures the commutation relation $[\hat{\psi}(\sx), \dot{\hat{\psi}}(\sx ')] = \delta(\sx -\sx ') + \text{image terms}$ and $\eta = \pm 1$ corresponds to a twisted (untwisted) field.

The Wightman function for the automorphic field between $\sx$ and $\sx '$ is then given by
\begin{equation}
    \begin{split}
        W_{R_0}^{(D)} (\sx, \sx ') &= \langle 0| \hat{\psi}_D(\sx) \hat{\psi}_D(\sx') |0\rangle
        \\
        &= \frac{1}{\mathcal{N}} \sum_{n,m=-\infty}^{\infty} \eta^n \eta^m \langle 0| \hat{\phi}(J^n_{0_D} \sx) \hat{\phi}(J^m_{0_D} \sx ') |0 \rangle
        \\
        &= \frac{1}{\mathcal{N}} \sum_{n,m=-\infty}^{\infty} \eta^n \eta^m W_\mathcal{R} (J^n_{0_D} \sx, J^m_{0_D} \sx ')\\
        &= \sum_{m=-\infty}^{\infty} \eta^m W_\mathcal{R} (\sx, J^m_{0_D} \sx '),
    \end{split}
    \label{eq:Wightman_D_sum}
\end{equation}
where $J^n_{0_D}: (\tau, r, y, z) \mapsto (\tau, r, y, z  + n l_D)$ \cite{identification_1, identification_2} 
generates a cylindrical spacetime with characteristic circumference $l_D$. 
We use the subscript $D$ here, in anticipation of the next section, where we shall superpose two distinct lengths $l_A$ and $l_B$.

More explicitly, along a uniformly accelerated trajectory, the Wightman function for a single identified spacetime is given by
 \begin{equation}
    \begin{split}
        W_{R_0}^{(D)} (\sx, \sx ') =& \frac{1}{\mathcal{N}} \sum_{n,m} \eta^n \eta^m \left[ \frac{\text{sgn}(\tau-\tau') \delta (\sigma _\mathcal{R}(J^n_{0_D} \sx, J^m_{0_D} \sx '))}{4\pi i}\right.\\&\left.- \frac{1}{4 \pi ^2 \sigma _\mathcal{R} (J^n_{0_D} \sx, J^m_{0_D} \sx ')}\right] ,
    \end{split}
    \label{eq:Wightman_D}
\end{equation}
where the geodesic distance squared with $l_D$-compactification orthogonal to the accelerated trajectory is given by
\begin{equation}
    \sigma _\mathcal{R}(J_{0_D}^n\sx, J_{0_D}^m \sx') = \frac{4}{a^2}\sinh^2\left[\frac{a}{2}(\tau-\tau')\right]-l_D^2(n-m)^2 .
    \label{eq:geodesic distance_D}
\end{equation}

The Wightman function in Eq. \eqref{eq:Wightman_D} can be expressed as a single sum akin to the last line of  Eq. \eqref{eq:Wightman_D_sum}.
However, we note that the double sum expression used in Eq. \eqref{eq:Wightman_D} allows for a more explicit contrast to the Wightman function of the automorphic fields associated to the spacetimes in a superposition as described by 
\cite{EduAlexDani2016}
\begin{equation}
    \begin{split}
        W_{R_0}^{(AB)}(\sx, \sx') &= \langle 0| \hat{\psi}_A(\sx) \hat{\psi}_B(\sx') |0\rangle
        \\
        &= \frac{1}{\mathcal{N}}\sum_{n,m=-\infty}^{\infty} \eta^n \eta^m \langle 0| \hat{\phi}(J^n_{0_A} \sx) \hat{\phi}(J^m_{0_B} \sx ') |0 \rangle
        \\
        &= \frac{1}{\mathcal{N}} \sum_{n,m=-\infty}^{\infty} \eta^n \eta^m W_\mathcal{R} (J^n_{0_A} \sx, J^m_{0_B} \sx '),
    \end{split}
    \label{eq:Wightman_cross_sum}
\end{equation}
or more explicitly as
\begin{equation}
    \begin{split}
        W_{R_0}^{(AB)} (\sx, \sx ') =& \frac{1}{\mathcal{N}} \sum_{n,m} \eta^n \eta^m \left[ \frac{\text{sgn}(\tau-\tau') \delta (\sigma _\mathcal{R}(J^n_{0_A} \sx, J^m_{0_B} \sx '))}{4\pi i}\right.\\&\left.- \frac{1}{4 \pi ^2 \sigma _\mathcal{R}(J^n_{0_A} \sx, J^m_{0_B} \sx ')}\right] ,
    \end{split}
    \label{eq:Wightman_AB}
\end{equation}
where the geodesic distance squared is now 
\begin{equation}
    \sigma _\mathcal{R}(J_{0_A}^n\sx, J_{0_B}^m \sx') = \frac{4}{a^2}\sinh^2\left[\frac{a}{2}(\tau-\tau')\right]-(l_A n-l_B m)^2 .
    \label{eq:geodesic distance_AB}
\end{equation}

\section{Detector Probe}\label{sec:detector}

In this section, we present the UDW detector model and how it can serve as an operational probe of spacetime when prepared in a superposed state.
We will also provide a more formal introduction to the Unruh effect.

\subsection{Quantum-Controlled Unruh-DeWitt Model}
\label{sec:UDW-model}

To study the effects of spacetime superposition on the behavior of an accelerated detector, we attribute a Hilbert space to the spacetime in addition to the field and to the detector \cite{FAZM_2022_BTZ}. 
In particular, such a system is described by a Hilbert space $\mathcal{H} = \mathcal{H}_{R_0} \otimes \mathcal{H}_F \otimes \mathcal{H}_{\text{UDW}}$ where
$\mathcal{H}_{\text{UDW}}$ is the Hilbert space of states for a two-level (UDW) detector; $\mathcal{H}_F$ is the Hilbert space of the quantum field's states; and
$\mathcal{H}_{R_0}$ is the Hilbert space spanned by the states of the spacetime being identified with characteristic circumference $l_D \in \mathbb{R}$
\footnote{$\mathcal{H}_{R_0}$ is the Hilbert space spanned by $\{|l_D\rangle\}_{l_D \in \mathbb{R}}$, where $|l_D\rangle$ satisfies $\hat{O}_{\text{QG}} |l_D\rangle = l_D |l_D\rangle$. Akin to the continuous position operator in standard quantum mechanics, orthonormality is defined using the Dirac delta function. Some describe this as \emph{rigged} Hilbert space \cite{Madrid_2005_Rigged-Hilbert-space,Madrid_2001_PhDThesis}.}.

In order for us to operationally access information arising from a spacetime superposition, we consider a UDW detector with a control degree of freedom associated to the spacetime. The initial state of the system can then be expressed as
\begin{equation}
    |\psi \rangle = \frac{1}{\sqrt{2}}\left( |l_A\rangle + |l_B\rangle \right) \otimes |0\rangle_\phi \otimes |g\rangle,
    \label{eq:Inital_state_of_system}
\end{equation}
where $\ket{g}$ is the initial state of the UDW detector, ground (excited) state when $\Omega > 0$ $(\Omega < 0)$, while $\ket{0}_\phi$ is the vacuum of the field $\hat{\phi}$ in an infinite Minkowski spacetime.

The interaction Hamiltonian describing the coupling among the detector, the quantum field, and the spacetime is \cite{FOZ_2020_suptrajUDW, FOMZ_2021, FMZ_2021, BCRAB_2020_Unruh-in-superposition, qts_qft} 
\begin{equation}
    \hat{H}_{\text{int.}} (\tau) = \lambda \,\chi(\tau) \sum_{D=A,B}  |l_D\rangle \langle l_D| \otimes \hat{\psi}_D(\sx(\tau)) \otimes \hat{m}(\tau),
    \label{eq:Interaction Hamiltonian}
\end{equation}
where $\lambda \ll 1$ is the coupling constant, $\tau$ is the proper time coordinate in the detector's reference frame, and $\hat{m}(\tau) = |e\rangle \langle g| e^{i\Omega \tau} + |g\rangle \langle e| e^{-i\Omega \tau}$ represents the detector's monopole operator
triggering transitions between the internal energy eigenstates $|g\rangle, |e\rangle$ with energy gap $\Omega$. 
Meanwhile, $\hat{\psi}_D(\sx)$ is the field operator \eqref{eq:qft automorphic field} for the $l_D$-compactified spacetime, computed along the trajectory of the detector $\sx(\tau)$.  
The time dependent switching function $\chi(\tau)$ gradually turns on and off the  interaction of the detector with the field.
We shall take it to be
\begin{equation}
\chi(\tau) = \exp\left(-\frac{\tau^2}{2\sigma^2}\right), 
    \label{eq:Gaussian switching function}
\end{equation}
where $\sigma$ is the characteristic interaction time.

\subsection{Detector Dynamics}\label{sec:detector-dynamics}

The perturbative time-evolution operator, in the interaction picture, is given by
\begin{equation}
    \begin{split}
        \hat{U} &= \mathcal{T} \exp\left( - i \int_{-\infty}^\infty d\tau \hat{H}_{\text{int.}}(\tau)\right)\\
        &= I - i\lambda \int_{-\infty}^\infty d\tau \hat{H}_{\text{int.}}(\tau)\\
        &- \lambda^2 \int_{-\infty}^\infty d\tau \int_{-\infty}^\tau d\tau' \hat{H}_{\text{int.}}(\tau)\hat{H}_{\text{int.}}(\tau') +\mathcal{O}(\lambda^3),
    \end{split}
    \label{eq:time evolve operator}
\end{equation}
where $\mathcal{T}$ denotes time-ordering.
We expand the time-evolution operator perturbatively in the Dyson series up to second order since we assume a weak interaction $(\lambda \ll 1)$. We evolve the initial state over time using
\begin{equation}
    \hat{U} |\psi \rangle  = \frac{1}{\sqrt{2}}\left( \hat{U}_A|l_A\rangle + \hat{U}_B|l_B\rangle\right)\otimes|0\rangle \otimes |g\rangle,
\label{eq:state evolve}    
\end{equation}
where $\hat{U}_D$ is the time-evolution operator generated by the $D$ term of the interaction Hamiltonian\footnote{Even though $\hat{U}_D$ is 2nd order in Hamiltonian, the cross-terms do not contribute to the density matrix due to the orthogonality of $\ket{l_A}$ and $\ket{l_B}$.} \eqref{eq:Interaction Hamiltonian}.

After the interaction, if we trace over the  the  field states \emph{and} over the control, we obtain 
\begin{equation}
    P^{(\text{Tr})}_E = \frac{\lambda^2}{2}(P_A+P_B),
    \label{eq:trace out transition probability}
\end{equation}
which is just the classical mixture of individual contributions from spacetimes with circumferences $l_A$ and $l_B$ \cite{FAZM_2023_flat,EduAlexDani2016}.
However, under the assumption that we can conditionally measure the control state in the superposition basis $|\pm\rangle = (|l_A\rangle\pm|l_B\rangle)/\sqrt{2}$ we obtain the final conditional state of the detector (not normalized)  
\begin{equation}
    \hat{\rho}_D^{(\pm)} = 
    \begin{pmatrix}
        P_G^{(\pm)} & 0 \\
        0 & P_E^{(\pm)}
    \end{pmatrix},
    \label{eq:final conditional state}
\end{equation}
after tracing over the field states,   
where $P_E^{(\pm)} \equiv P^{(\pm)}(g\to e)$ is the transition probability of the detector measured in the corresponding $|\pm\rangle$ state. This is given by \cite{FAZM_2023_flat}
\begin{equation}
    P_E^{(\pm)} = \frac{\lambda^2}{4}(P_A+P_B\pm 2L_{AB}),
    \label{eq:total transition probability}
\end{equation}
where
\begin{equation}
    P_D = \int_{-\infty}^\infty d\tau \int_{-\infty}^\infty d\tau' \, \mu(\tau) \mu^*(\tau') \, W_{R_0}^{(D)} (\sx, \sx '),
    \label{eq:transition probability single detector}
\end{equation}
is the response function of the detector travelling along a uniformly accelerated trajectory in a single identified spacetime with characteristic circumference $l_D$  with
\begin{equation}
    \mu(\tau) = \chi(\tau) e^{-i\Omega\tau} = \exp\left(-\frac{\tau^2}{2\sigma^2}\right) e^{-i\Omega\tau}. 
\end{equation}

Also appearing in the conditional transition probability is a cross-correlation term
\begin{equation}
    L_{AB} = \int_{-\infty}^\infty d\tau \int_{-\infty}^\infty d\tau' \, \mu(\tau) \mu^*(\tau') \, W_{R_0}^{(AB)} (\sx, \sx '),
    \label{eq:cross-correlation term}
\end{equation}
evaluated with respect to the field quantized on both spacetimes with characteristic circumferences $l_A$ and $l_B$. This quantifies the quantum interference between the two spacetimes. 
Note that as $\sigma\to\infty$, the switching function allows the detector to interact with the field for an arbitrarily long time. In a single spacetime this allows the detector to thermalize.

We can explicitly obtain the ``local'' contribution   (\ref{eq:transition probability single detector}) and the cross-correlation term (\ref{eq:cross-correlation term}) using the Wightman functions (\ref{eq:Wightman_D}, \ref{eq:Wightman_AB}) to get
\begin{widetext} 
\begin{align}
    \begin{split}
        P_D = \frac{\sqrt{\pi}\sigma}{\mathcal{N}}&\left[\, \, \,\sum_{\substack{n,m=-\infty\\n=m}}^\infty\left( -\frac{\Omega}{4\pi} - \frac{a^2}{8\pi^2} \int_0^\infty ds \left( \frac{e^{-\frac{s^2}{4\sigma^2}} \cos(\Omega s)}{\sinh^2(\frac{as}{2})}-\frac{4}{a^2s^2}\right) \right)\right.\\
        & - \left.\sum_{\substack{n,m=-\infty\\n>m}}^\infty \frac{1}{2\pi k_D} \frac{e^\frac{- (\sinh^{-1}(ak_D/2))^2}{a^2\sigma^2}}{\sqrt{\frac{a^2}{4}k_D^2 + 1}} \sin(\frac{2\Omega}{a} \sinh^{-1}\left(\frac{a}{2}k_D\right))\right.\\
        &+ \left. \sum_{\substack{n,m=-\infty\\ n\neq m}}^\infty \frac{\sigma}{4\pi\sqrt{\pi}k_D}\int_{-\infty}^{\infty} dz\, \, \frac{e^{-(\Omega-z)^2\sigma^2}\coth(\frac{\pi z}{a})}{\sqrt{\frac{a^2}{4}k_D^2+1}} \sin(\frac{2z}{a} \sinh^{-1}\left(\frac{a}{2}k_D\right))\right], 
        \label{eq:P_D form} 
    \end{split}
    \\
    \begin{split}
        L_{AB} = \frac{\sqrt{\pi}\sigma}{\mathcal{N}}&\left[\, \, \,\sum_{\substack{n,m=-\infty\\l_A n=l_B m}}^\infty\left( -\frac{\Omega}{4\pi} - \frac{a^2}{8\pi^2} \int_0^\infty ds \left( \frac{e^{-\frac{s^2}{4\sigma^2}} \cos(\Omega s)}{\sinh^2(\frac{as}{2})}-\frac{4}{a^2s^2}\right) \right)\right.\\
        & - \left.\sum_{\substack{n,m=-\infty\\l_A n>l_B m}}^\infty \frac{1}{2\pi k_{AB}} \frac{e^\frac{- (\sinh^{-1}(a k_{AB}/2))^2}{a^2\sigma^2}}{\sqrt{\frac{a^2}{4}k_{AB}^2 + 1}} \sin(\frac{2\Omega}{a} \sinh^{-1}\left(\frac{a}{2}k_{AB}\right))\right.\\
        &+ \left. \sum_{\substack{n,m=-\infty\\l_A n\neq l_B m}}^\infty \frac{\sigma}{4\pi\sqrt{\pi} k_{AB}}\int_{-\infty}^{\infty} dz\, \, \frac{e^{-(\Omega-z)^2\sigma^2}\coth(\frac{\pi z}{a})}{\sqrt{\frac{a^2}{4}k_{AB}^2+1}} \sin(\frac{2z}{a} \sinh^{-1}\left(\frac{a}{2}k_{AB}\right))
        \right], \label{eq:L_AB form} 
    \end{split}
\end{align}
where $k_D = l_D(n-m)$, $k_{AB} = l_A n-l_B m$, $\mathcal{N} = \sum_n \eta^{2n}$, and $s = \tau - \tau'$ (the proper time difference).
\end{widetext}

Note that in the limit $a \to 0$, i.e., for a static detector, we retrieve the formulae for the superposition of cylindrically identified Minkowski spacetimes \cite{FAZM_2023_flat}, which we show in detail in \autoref{appendix:compare with a=0}. 

The first sum, conditioned on $n=m$, in (\ref{eq:P_D form}) corresponds to the transition probability of a single accelerated detector in Minkowski spacetime, which we call $P_\mathcal{R}$.
The cylindrical identification $R_0$ contributes to the remaining image sum terms. 
We computationally confirm in \autoref{appendix:convergence of image sum} that the sums in Eqs. (\ref{eq:P_D form}, \ref{eq:L_AB form}) converge with respect to the image sum terms.

\section{Results}\label{sec:results}

\begin{figure}[t]
    \centering
    \includegraphics[width=1\linewidth]{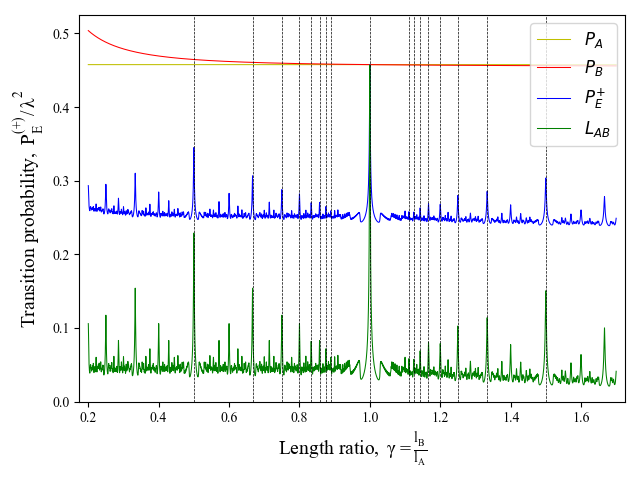}
    \caption{\justifying Plots of $P_A$ (yellow), $P_B$ (red), $P_E^{(+)}$  (blue) and
$L_{AB}$ (green),  
as a function of the ratio of the two characteristic circumferences $l_A \text{ and } l_B$ for an acceleration of $a\sigma = 10$ with energy gap $\Omega\sigma=0.01$. 
    Vertical grid-lines were added to emphasize the peaks at rational values (e.g., $3/2, 6/5, 1, 4/5, 1/2$).  $\sigma$ is the characteristic interaction time.
    }
    \label{fig:pe-length-ratio}
\end{figure}

In \autoref{fig:pe-length-ratio} we  plot  
$P_A, P_B, P_E^{(+)}$ and $L_{AB}$ 
as a function of 
$\gamma \equiv l_B/l_A$ for a detector acceleration of $a\sigma = 10$ and gap
$\Omega\sigma = 0.01$.
Akin to previous work examining UDW detectors in superposed spacetimes \cite{FAZM_2022_BTZ,FAZM_2023_flat,Diki_2023_BTZ_rotate}, we observe resonant peaks in the transition probability at rational values of $\gamma$. Some of these are highlighted in the figure using vertical lines (e.g., at $\gamma =$ 3/2, 6/5, 1, 4/5, 1/2). We expect a countably infinite number of these discrete resonant peaks, with one arising from every rational value of $\gamma$.

The resonant nature of the plots can be attributed to the $k_{AB}=l_A n-l_B m$ terms in Eq. (\ref{eq:L_AB form}). 
We observe ``continuous'' resonance peaks commensurate with previous studies  \cite{Diki_2023_BTZ_rotate,FAZM_2022_BTZ}, but
not the ``discontinuous'' resonance peaks that were previously observed for a static detector in cylindrically identified superposed Minkowski spacetime \cite{FAZM_2023_flat}. 
In the limit $a\sigma\to0$ (see \autoref{appendix:compare with a=0}), our resonance peaks remained ``continuous'', suggesting a numerical discrepancy with Figure 1 of Ref. \cite{FAZM_2023_flat}, though these would not change the overarching results, which we corroborate in this 
paper\footnote{Indeed, we were able to confirm that the code used for \cite{FAZM_2023_flat} uses different numbers of terms when computing the curves on and off resonance. When correcting for this, one recovers the ``continuous'' results, as we did in \cite{Goel_Accelerated_Detector_in}.}.  From a physical perspective, we expect resonances to have a finite height and width because of the finite observation time.

\begin{figure}[t]
    \centering
    \includegraphics[width=1\linewidth]{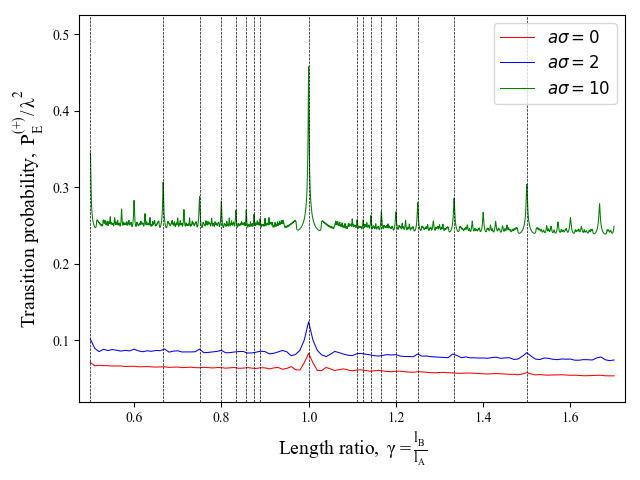}
    \caption{\justifying Plots of $P_E^{(+)}$ as a function of the ratio of the two characteristic circumferences $l_A \text{ and } l_B$ for three different accelerations $a\sigma = 0, 2, 10$ 
    and fixed $\Omega\sigma=0.01$.  
    Vertical grid-lines were added for distinctive peaks at rational values (e.g., $3/2, 6/5, 1, 4/5, 1/2$). $\sigma$ is the characteristic interaction time. }
    \label{fig:pe-gamma-different-acc}
\end{figure}

The effect of acceleration on $P_E^{(+)}$ is further described in \autoref{fig:pe-gamma-different-acc}, where we show that the resonance peaks are more pronounced for increasing acceleration.
When $a\sigma = 10$, the peaks resemble those seen in the BTZ black hole cases \cite{Diki_2023_BTZ_rotate, FAZM_2022_BTZ} whereas the static detector case $a=0$, corroborates the results from \cite{FAZM_2023_flat} (except for their discontinuous resonance peaks).
A more detailed account of the static detector setup can be found in \autoref{appendix:compare with a=0}.

\begin{figure}
    \centering
    \begin{subfigure}{0.9\linewidth}
        \centering
        \includegraphics[width=1\linewidth]{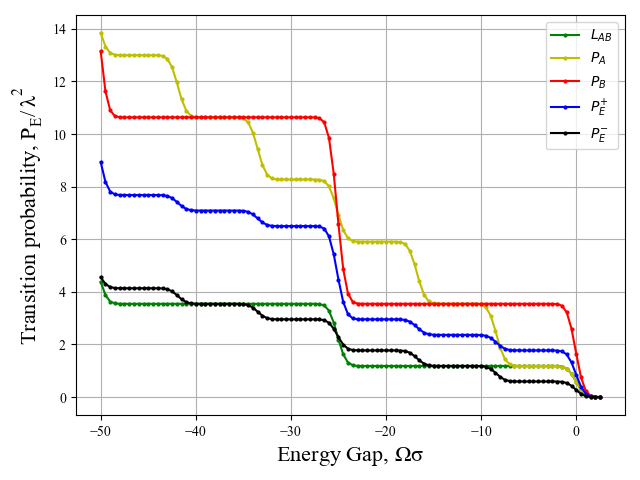}
        \caption{Acceleration $a\sigma = 0$}
        \label{fig:pe-energy-gap:a}
    \end{subfigure}
    \begin{subfigure}{0.9\linewidth}
        \centering
        \includegraphics[width=1\linewidth]{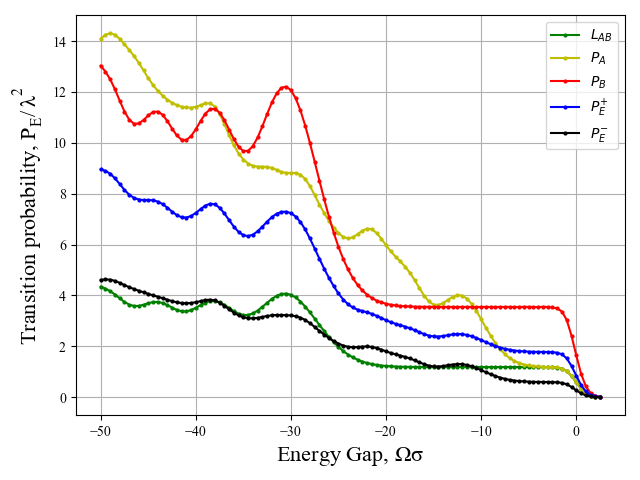}
        \caption{Acceleration $a\sigma = 2$}
        \label{fig:pe-energy-gap:b}
    \end{subfigure}
    % \hfill
    \begin{subfigure}{0.9\linewidth}
        \centering
        \includegraphics[width=1\linewidth]{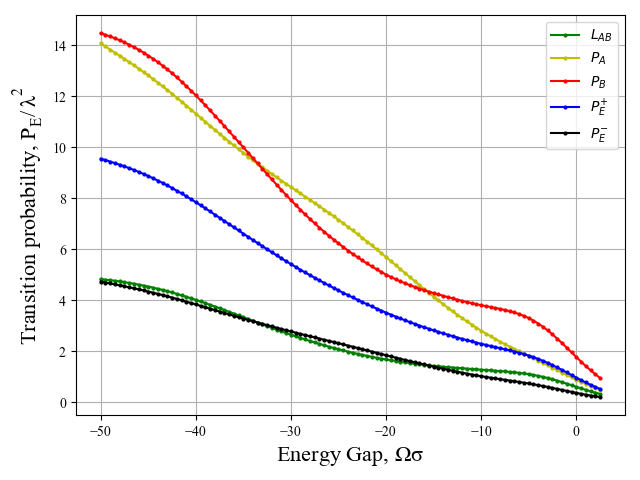}
        \caption{Acceleration $a\sigma = 15$}
        \label{fig:pe-energy-gap:c}
    \end{subfigure}
    \caption{\justifying Plots of $P_A$ (yellow), $P_B$ (red), $P_E^{(+)}$  (blue), $P_E^{(-)}$  (black),  and
$L_{AB}$ (green),
  as a function of the energy gap $\Omega\sigma$. The parameters $l_A/\sigma = 0.75, l_B/\sigma=0.25$. Accelerations are $a\sigma = 0, 2$, and $15$, from top to bottom. $\sigma$ is the characteristic interaction time.
    }
    \label{fig:pe-energy-gap}
\end{figure}

\begin{figure*}[t]
    \centering
    \begin{subfigure}{0.475\linewidth}
        \centering
        \includegraphics[width=\textwidth]{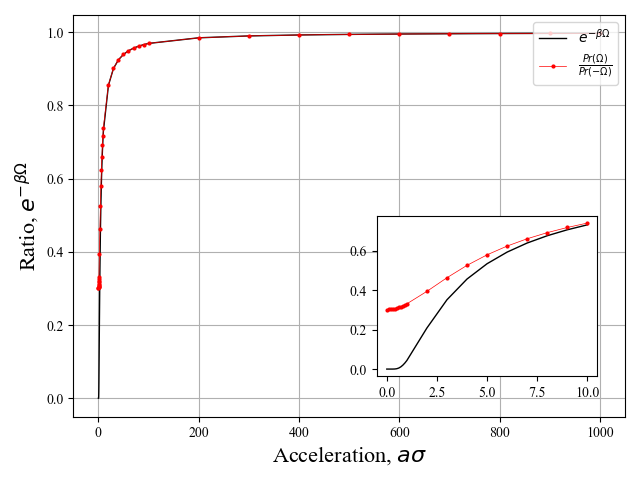}
        \caption{Low energy gap $|\Omega\sigma|=0.5$}
        \label{fig:EDR_DBC_a}
    \end{subfigure}
    \hfill
    \begin{subfigure}{0.475\linewidth}
        \centering
        \includegraphics[width=\textwidth]{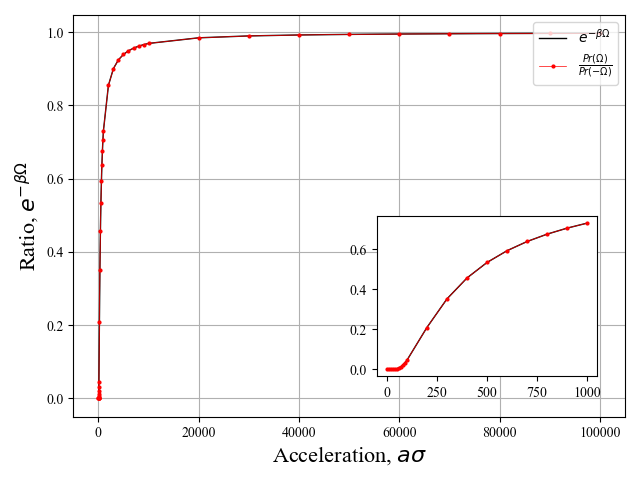}
        \caption{High energy gap $|\Omega\sigma|=50$}
        \label{fig:EDR_DBC_b}
    \end{subfigure}
    \caption{\justifying Plot of the  EDR (red) and the thermal DBC $e^{-\beta\Omega}$ (black) as a function of acceleration, $a \sigma$ for fixed energy gaps $\Omega\sigma = 0.5, 50$ with parameters $l_A/\sigma = 0.75, ~l_B/\sigma=0.25$, characterizing the compactification lengths of the superposed spacetimes. Inset plots show the behaviour at low acceleration. We note that the thermal behavior is recovered if the characteristic interaction time, $\sigma$, is longer than $\Omega^{-1}$ or $a^{-1}$.}
\label{fig:EDR_DBC}
\end{figure*}

The similarities between our results for an accelerated detector and those of a static detector in BTZ spacetimes (in spite of the difference in the number of spatial dimensions) can be understood through the lens of the equivalence principle: the near-horizon geometry of a black hole is the same as the Rindler geometry, seen by an accelerated observer in Minkowski spacetime. In addition to the resonance peaks becoming more pronounced, as acceleration increases, the detector experiences a higher Unruh temperature, 
which allows for an easier transition from $|g\rangle \to |e\rangle$ and thus the transition probability $P_E^{(+)}$ increases.

To further understand how acceleration affects the transition probability, we depict in \autoref{fig:pe-energy-gap} the various quantities of interest as a function of the energy gap for different accelerations. Recall that a negative energy gap (e.g., $\Omega\sigma = -30 $) means that the detector is initialized in the excited state and deexcites to the ground state. 

In Figure \ref{fig:pe-energy-gap:a}, we recover the results for a static detector ($a\sigma=0$) in cylindrically identified Minkowski spacetime, where we observe oscillations about the values in flat Minkowski spacetime \cite{EduAlexDani2016}. These oscillations resemble a descending staircase, where the frequency of the steps is proportional to the characteristic length ratio and inversely proportional to the length of a given step. 

In the case of an accelerating detector, we observe additional fluctuations from these oscillations, which are more pronounced for larger (negative) energy gaps. 
Interestingly, we see that in the limit of large acceleration the ``local'' contributions to the transition probabilities $P_A$, $P_B$ appear to tend towards a common curve.
This can be understood by observing that as $a\to\infty$, the transition probability given in Eq. (26) has no dependence on $k_D$; instead, $\lim_{a\to\infty}P_D= P_\mathcal{R}$. Physically, this means that the detector response at high Unruh temperatures, or high accelerations, $a l_D \gg 1$ is no longer sensitive to the compactification of space, i.e., thermal wavelength is much shorter than the box size.

For positive energy gaps, we note that the larger acceleration leads to an increase in all quantities, which can be seen by comparing the $\Omega\sigma>0$ regions of \autoref{fig:pe-energy-gap:b} and \autoref{fig:pe-energy-gap:c}.
This is in agreement with previous work \cite{Lee_2013_PhDThesis} and is commensurate with the increase in $P_E^{(+)}$ in \autoref{fig:pe-gamma-different-acc}.

Finally, in \autoref{fig:EDR_DBC} we illustrate
the thermalization of our detector, showing the relationship between the EDR, $P_E^{(+)}(\Omega)/P^{(+)}_E(-\Omega)$, and the DBC of the detector when the control is measured in the $\ket{+}$ space-time superposition state. 

While we note the discrepancy between the EDR and the DBC in \autoref{fig:EDR_DBC_a} for low $a\sigma$ and small energy gap, thermalization is a late-time behaviour corresponding to long interaction times, $\sigma$, compared to the thermal time scale, $a^{-1}$. 
In the limit of large $a\sigma$, we observe that the EDR indeed satisfies the DBC. In this sense, we conclude that a detector undergoing uniform acceleration in a superposition of spacetimes  thermalize if they share the same thermal state.

This behaviour might appear to be somewhat surprising at first glance given that, in general, a UDW detector in a superposition of trajectories with acceleration $a$ does not thermalize \cite{FOZ_2020_suptrajUDW}. This discrepancy can be attributed to the fact that the trajectories considered in \cite{FOZ_2020_suptrajUDW} that have the same acceleration $a$ do not share a common Rindler horizon. Instead, our work reinforces the conclusions of \cite{FooZych_2023_S-of-thermalisation}, that the presence of a shared Rindler horizon, and thus a shared thermal state, is key to the thermalization of UDW detectors in superposition of trajectories and spacetimes.

\section{Conclusion}\label{sec:conclusion}

In this paper, we studied basic quantum-gravitational effects of spacetime superposition on an accelerating two-level UDW detector coupled to a massless scalar quantum field. 
Each branch of the superposition is a cylindrically identified Minkowski spacetime 
with a distinct identification length. 
In particular, we have shown that the UDW detector's acceleration enhances the resonance peaks at rational values of the ratio $\gamma$ of the characteristic lengths. Acceleration also increases the overall amplitude of the transition probability.

In the limit of small acceleration, we recover the previous results for the static case \cite{FAZM_2023_flat}, though we show that even in this scenario, the resonance peaks appear to arise in a ``continuous'' fashion.
By revisiting the previous claim of ``discontinuous'' resonances, we suggest that the resonance peaks arising from spacetime superpositions might be ``continuous'' in general. This would be in accord  with the resonance peaks in superposed BTZ spacetimes \cite{FAZM_2022_BTZ, Diki_2023_BTZ_rotate}. 

Acceleration also blurs the sharpness of the steps (or oscillations) of the detector response, the blur increasing for higher accelerations,   as shown in \autoref{fig:pe-energy-gap}. 
As the detector accelerates, it becomes less sensitive to the spacetime identification's characteristic length in that the two ``local'' contributions of the transition probabilities converge.

Finally,  we found that a detector undergoing uniform acceleration in a quantum superposition of spacetimes   does indeed thermalize,  in the sense that the EDR satisfies the DBC condition, at the Unruh temperature, for sufficiently long interaction time.

Although the Unruh effect has been studied for detectors traveling along a superposition of accelerating trajectories \cite{FOZ_2020_suptrajUDW,BCRAB_2020_Unruh-in-superposition,FooZych_2023_S-of-thermalisation}, 
our results constitute the first study of accelerating detectors in a genuine superposition of spacetimes.
Moreover, this paper also provides the first study of thermal response for a detector in a superposition of (3+1)-dimensional spacetimes.

A natural next step would be a study of the response of accelerating detectors in a superposition of Minkowski spacetimes with cylindrical identification in two directions. This might offer insight into more general notions of area quantization and their relation to concepts such as entropy.

\section*{Acknowledgment}
This work was supported in part by the Natural Sciences and Engineering Research Council of Canada, and Perimeter Institute for Theoretical Physics.
Research at Perimeter Institute is supported in part by the Government of Canada through the Department of Innovation, Science and Economic Development and by the Province of Ontario through the Ministry of Colleges and Universities.
We would like to thank Joshua Foo and Cemile Arabaci for helpful correspondence and sharing some of the code used in \cite{FAZM_2023_flat} with us. 
E. A. P. acknowledges support from the Province of Ontario as a recipient of an Ontario Graduate Scholarship, and from NSERC as a recipient of a CGS-D Scholarship.
M. R. P.-R. gratefully acknowledges the support from the Mike and Ophelia Lazaridis Graduate Fellowship. M. T. acknowledges Canadian Institute for Theoretical Astrophysics (CITA) for financial support.

\newpage

\appendix
\begin{widetext}

\section{Specifying the UDW detector model}

\subsection{UDW detector in superposed spacetime}

The Hilbert space for our joint system is given by $\mathcal{H} = \mathcal{H}_{R_0} \otimes \mathcal{H}_F \otimes \mathcal{H}_{UDW}$ which is a tensor product of the Hilbert spaces corresponding to the quantum spacetime, the quantum field, and the matter degrees of freedom (i.e., the detector) respectively.

Consider the cylindrically identified Minkowski spacetime $\mathcal{M}_0$ in a superposition of two characteristic lengths $l_A$ and $l_B$. 
Further suppose that the quantum field is in the vacuum state. 
Quantum matter is modeled by an Unruh-DeWitt detector assumed to be initialized in the ground state. 

The initial state of the total system is then given by
\begin{equation}
    \ket{\psi} = \frac{1}{\sqrt{2}} (\ket{l_A}+\ket{l_B})\ket{0}\ket{g},
\end{equation}
where the spacetime superposition will serve as the quantum control degree of freedom for our eventual detector measurement.

The interaction Hamiltonian in the interaction picture
\begin{equation}
    \hat{H}_{\text{int}} = \lambda~\eta(\tau)\hat{\sigma}(\tau) \sum_{D=A,B} \hat{\psi}_D(\sx)\otimes\ketbra{l_D}{l_D}
\end{equation}
(note the order of the Hilbert space has been reversed) defines the coupling and requires a common time coordinate. There,
\begin{equation}
    \hat{\sigma}(\tau) = \ketbra{e}{g} e^{\ii\Omega \tau} + \ketbra{g}{e} e^{-\ii\Omega \tau}
\end{equation}
is the time-dependent monopole operator of the detector, obtained via $\hat{\sigma}_{x}(\tau)=e^{\ii\hat{\mathfrak{h}}\tau}\hat{\sigma}_{x}e^{-\ii\hat{\mathfrak{h}}\tau}$, where $\hat{\mathfrak{h}}=\frac{\Omega}{2}(\hat{\sigma}_{z}+\openone)$ is the detector's free Hamiltonian, with $\sigma_z$ and $\sigma_{x}$ the standard Pauli matrices.

 The time-evolution operator is given in the interaction picture by the time-ordered exponential of the Hamiltonian, 
\begin{equation}
    \hat{U} = \hat{\mathcal{T}} \exp \left( - \ii \int_{-\infty}^{\infty} \dd \tau \: \hat{H}_\mathrm{int}(\tau) \right) 
\end{equation}
which can be expanded perturbatively as a Dyson series expansion:
\begin{align}
    \hat{U} &= \mathds{1} + \hat{U}^{(1)} + \hat{U}^{(2)} + \mathcal{O}(\lambda^3)
    \nonumber 
    \\
    &= \mathds{1} - \ii \lambda \int_{-\infty}^{\infty} \dd \tau \: \hat{H}_\mathrm{int}(\tau) 
     - \lambda^2 \int_{-\infty}^{\infty} \dd \tau \int_{-\infty}^\tau \dd \tau' \hat{H}_\mathrm{int}(\tau) \hat{H}_\mathrm{int}(\tau') + \mathcal{O}(\lambda^3)
\end{align}
We evolve the initial state in time, trace over the field states and then measure the control state in the superposition basis $\ket{\pm}=(|l_A \rangle \pm |l_B \rangle)/\sqrt{2}$. 
Applying this procedure yields the final state of the detector, from which we can identify the conditional transition probability conditioned on the $\ket{\pm}$ basis.

\subsection{Final Density Matrix}

We will begin by identifying how the interaction Hamiltonian generates time evolution of our previously assumed initial state 
$$
\ket{\psi} = \frac{1}{\sqrt{2}} (\ket{l_A}+\ket{l_B})\ket{0}\ket{g}
$$
of the total system.

For the detector's (matter) Hilbert space, note that
\begin{equation}
    \hat{\sigma}(\tau)\ket{g} = (\ketbra{e}{g} e^{\ii\Omega \tau} + \ketbra{g}{e} e^{-\ii\Omega \tau})\ket{g} = e^{\ii\Omega \tau}\ket{e} + 0
\end{equation}
and
\begin{align}
    \hat{\sigma}(\tau)\hat{\sigma}(\tau')\ket{g} &= (\ketbra{e}{g} e^{\ii\Omega \tau} + \ketbra{g}{e} e^{-\ii\Omega \tau})(\ketbra{e}{g} e^{\ii\Omega \tau'} + \ketbra{g}{e} e^{-\ii\Omega \tau'})\ket{g}   \nonumber
    \\
    &= (\ketbra{e}{e} e^{\ii\Omega (\tau-\tau')} + \ketbra{g}{g} e^{-\ii\Omega (\tau-\tau')})\ket{g}    \nonumber
    \\
    &= e^{-\ii\Omega (\tau-\tau')}\ket{g}
\end{align}

Further note that (for the field and spacetime Hilbert spaces) we have
\begin{equation}
    \sum_{D=A,B} \hat{\psi}_D(\sx)\hat{\psi}_D(\sx')\otimes\ketbra{l_D}{l_D} \left( \frac{1}{\sqrt{2}} (\ket{l_A}+\ket{l_B})\ket{0} \right) = \frac{1}{\sqrt{2}} \sum_{D=A,B} \hat{\psi}_D(\sx)\hat{\psi}_D(\sx')\ket{0} \otimes \ket{l_D} 
\end{equation}
since
\begin{align}
    \sum_{D=A,B} \ketbra{l_D}{l_D} \left( \frac{1}{\sqrt{2}} (\ket{l_A}+\ket{l_B}) \right) &= \frac{1}{\sqrt{2}} (\ketbra{l_A}{l_A}+\ketbra{l_B}{l_B})(\ket{l_A}+\ket{l_B})
   \nonumber \\
    &= \frac{1}{\sqrt{2}} (\ket{l_A}+\ket{l_B})
\end{align}
provided that $\braket{l_A|l_B}=0$.

Applying the time-evolution (unitary) operator yields
\begin{align}
    \hat{U}\ket{\psi} &= \left(\mathds{1} + \hat{U}^{(1)} + \hat{U}^{(2)}\right) \frac{1}{\sqrt{2}} (\ket{l_A}+\ket{l_B})\ket{0}\ket{g} 
    \nonumber\\
    &= \left(\mathds{1} - \ii \lambda \int_{-\infty}^{\infty} \dd \tau \: \hat{H}_\mathrm{int}(\tau) - \lambda^2 \int_{-\infty}^{\infty} \dd \tau \int_{-\infty}^\tau \dd \tau' \hat{H}_\mathrm{int}(\tau) \hat{H}_\mathrm{int}(\tau')\right) \frac{1}{\sqrt{2}} (\ket{l_A}+\ket{l_B})\ket{0}\ket{g}
   \nonumber \\
    &=  \frac{1}{\sqrt{2}} (\ket{l_A}+\ket{l_B})\ket{0}\ket{g} - \ii \lambda \int_{-\infty}^{\infty} \dd \tau \: \hat{H}_\mathrm{int}(\tau)  \frac{1}{\sqrt{2}} (\ket{l_A}+\ket{l_B})\ket{0}\ket{g} \nonumber \\
    &\quad - \lambda^2 \int_{-\infty}^{\infty} \dd \tau \int_{-\infty}^\tau \dd \tau' \hat{H}_\mathrm{int}(\tau) \hat{H}_\mathrm{int}(\tau') \frac{1}{\sqrt{2}} (\ket{l_A}+\ket{l_B})\ket{0}\ket{g}
  \nonumber  \\
    &=  \frac{1}{\sqrt{2}} (\ket{l_A}+\ket{l_B})\ket{0}\ket{g} - \ii \lambda \int_{-\infty}^{\infty} \dd \tau \: \left(\eta(\tau)\hat{\sigma}(\tau) \sum_{D=A,B} \hat{\psi}_D(\sx)\otimes\ketbra{l_D}{l_D}\right) \frac{1}{\sqrt{2}} (\ket{l_A}+\ket{l_B})\ket{0}\ket{g} \nonumber \\
    &\quad - \lambda^2 \int_{-\infty}^{\infty} \dd \tau \int_{-\infty}^\tau \dd \tau' \left(\eta(\tau)\hat{\sigma}(\tau) \sum_{D=A,B} \hat{\psi}_D(\sx)\otimes\ketbra{l_D}{l_D}\right) \nonumber \\
    &\qquad \qquad \qquad \qquad \times \left(\eta(\tau')\hat{\sigma}(\tau') \sum_{D=A,B} \hat{\psi}_D(\sx')\otimes\ketbra{l_D}{l_D}\right) \frac{1}{\sqrt{2}} (\ket{l_A}+\ket{l_B})\ket{0}\ket{g}
   \nonumber \\
    &=  \frac{1}{\sqrt{2}} (\ket{l_A}+\ket{l_B})\ket{0}\ket{g} - \ii \lambda \int_{-\infty}^{\infty} \dd \tau \: \left(\eta(\tau)\hat{\sigma}(\tau) \sum_{D=A,B} \hat{\psi}_D(\sx)\otimes\ketbra{l_D}{l_D}\right) \frac{1}{\sqrt{2}} (\ket{l_A}+\ket{l_B})\ket{0}\ket{g} \nonumber \\
    &\quad - \lambda^2 \int_{-\infty}^{\infty} \dd \tau \int_{-\infty}^\tau \dd \tau' \left(\eta(\tau)\eta(\tau')\hat{\sigma}(\tau)\hat{\sigma}(\tau') \sum_{D=A,B} \hat{\psi}_D(\sx)\hat{\psi}_D(\sx')\otimes\ketbra{l_D}{l_D}\right) \frac{1}{\sqrt{2}} (\ket{l_A}+\ket{l_B})\ket{0}\ket{g}
   \nonumber \\
    &=  \frac{1}{\sqrt{2}} (\ket{l_A}+\ket{l_B})\ket{0}\ket{g} - \frac{\ii}{\sqrt{2}} \lambda \int_{-\infty}^{\infty} \dd \tau \: \left(\eta(\tau)\hat{\sigma}(\tau)\ket{g} \sum_{D=A,B} \hat{\psi}_D(\sx)\ket{0}\otimes\ket{l_D}\right)  \nonumber \\
    &\quad - \frac{\lambda^2}{\sqrt{2}} \int_{-\infty}^{\infty} \dd \tau \int_{-\infty}^\tau \dd \tau' \left(\eta(\tau)\eta(\tau')\hat{\sigma}(\tau)\hat{\sigma}(\tau')\ket{g}  \sum_{D=A,B} \hat{\psi}_D(\sx)\hat{\psi}_D(\sx')\ket{0}\otimes\ket{l_D}\right)
   \nonumber \\
    &=  \frac{1}{\sqrt{2}} (\ket{l_A}+\ket{l_B})\ket{0}\ket{g} - \frac{\ii}{\sqrt{2}} \lambda \int_{-\infty}^{\infty} \dd \tau \: \left(\eta(\tau) e^{\ii\Omega \tau}\ket{e} \sum_{D=A,B} \hat{\psi}_D(\sx)\ket{0}\otimes\ket{l_D}\right)  \nonumber \\
    &\quad - \frac{\lambda^2}{\sqrt{2}} \int_{-\infty}^{\infty} \dd \tau \int_{-\infty}^\tau \dd \tau' \left(\eta(\tau)\eta(\tau') e^{-\ii\Omega (\tau-\tau')}\ket{g}  \sum_{D=A,B} \hat{\psi}_D(\sx)\hat{\psi}_D(\sx')\ket{0}\otimes\ket{l_D}\right)
\end{align}

We can now condition on the $\ket{\pm}=\frac{1}{\sqrt{2}}(\ket{l_A}\pm\ket{l_B})$ basis, we get
\begin{align}
    \bra{\pm}\hat{U}\ket{\psi} &=  \frac{1}{2}(\braket{l_A|l_A}\pm\braket{l_B|l_B})\ket{0}\ket{g} - \frac{\ii}{2} \lambda \int_{-\infty}^{\infty} \dd \tau \: \eta(\tau) e^{\ii\Omega \tau}\ket{e} \left( \hat{\psi}_A(\sx) \pm \hat{\psi}_B(\sx) \right) \ket{0}  \nonumber \\
    &\quad - \frac{\lambda^2}{2} \int_{-\infty}^{\infty} \dd \tau \int_{-\infty}^\tau \dd \tau' \eta(\tau)\eta(\tau') e^{-\ii\Omega (\tau-\tau')}\ket{g} \left( \hat{\psi}_A(\sx)\hat{\psi}_A(\sx') \pm \hat{\psi}_B(\sx)\hat{\psi}_B(\sx')\right)\ket{0}
\end{align}
since
\begin{equation}
    \bra{\pm} \frac{1}{\sqrt{2}} \sum_{D=A,B}\hat{O}_D\ket{0}\otimes\ket{l_D} = \frac{1}{2} \left(\hat{O}_A\otimes\braket{l_A|l_A} \pm \hat{O}_B\otimes\braket{l_B|l_B}\right)\ket{0}
\end{equation}
and
\begin{equation}
    \braket{l_A|l_A}\pm\braket{l_B|l_B}= 1\pm1.
\end{equation}

\bigskip

The next step is tracing out the field. Restricting ourselves to second order in the coupling, and noting that the odd-point functions vanish, we obtain
\begin{align}
    \tr_{\hat{\psi}}&\left( \bra{\pm}\hat{U}^{(0)}\ket{\psi} \bra{\psi}\hat{U}^{(2)\dagger}\ket{\pm} \right) 
   \nonumber \\
    &= \tr_{\hat{\psi}} \left( \frac{(1\pm1)}{2}\ket{0}\ket{g} (-\frac{\lambda^2}{2}) \int_{-\infty}^{\infty} \dd \tau \int_{-\infty}^\tau \dd \tau' \eta(\tau)\eta(\tau') e^{\ii\Omega (\tau-\tau')}\bra{g} \bra{0}\left( \hat{\psi}_A(\sx')\hat{\psi}_A(\sx) \pm \hat{\psi}_B(\sx')\hat{\psi}_B(\sx)\right) \right) 
   \nonumber \\
    &= \tr_{\hat{\psi}} \left( \frac{-\lambda^2}{4}(1\pm1)\ketbra{g}{g}  \int_{-\infty}^{\infty} \dd \tau \int_{-\infty}^\tau \dd \tau' \eta(\tau)\eta(\tau') e^{\ii\Omega (\tau-\tau')} \bra{0}\left( \hat{\psi}_A(\sx')\hat{\psi}_A(\sx) \pm \hat{\psi}_B(\sx')\hat{\psi}_B(\sx)\right)\ket{0} \right) 
   \nonumber \\
    &= - \frac{\lambda^2(1\pm1)}{4}\ketbra{g}{g} \int_{-\infty}^{\infty} \dd \tau \int_{-\infty}^\tau \dd \tau' \eta(\tau)\eta(\tau') e^{\ii\Omega (\tau-\tau')}\left( W(\sx_A',\sx_A) + W(\sx_B',\sx_B)\right)
\end{align}
Following similar steps yields
\begin{align}
    \tr_{\hat{\psi}}\left( \bra{\pm}\hat{U}^{(0)}\ket{\psi} \bra{\psi}\hat{U}^{(0)\dagger}\ket{\pm} \right) &= \frac14(2\pm2)\ket{g}\bra{g}
    \\
    \tr_{\hat{\psi}}\left( \bra{\pm}\hat{U}^{(0)}\ket{\psi} \bra{\psi}\hat{U}^{(2)\dagger}\ket{\pm} \right) &= 
    - \frac{\lambda^2(1\pm1)}{4}\ketbra{g}{g} \int_{-\infty}^{\infty} \dd \tau \int_{-\infty}^\tau \dd \tau' \eta(\tau)\eta(\tau') e^{\ii\Omega (\tau-\tau')}\left( W(\sx_A',\sx_A) + W(\sx_B',\sx_B)\right)
    \\
    \tr_{\hat{\psi}}\left( \bra{\pm}\hat{U}^{(2)}\ket{\psi} \bra{\psi}\hat{U}^{(0)\dagger}\ket{\pm} \right) &= - \frac{\lambda^2(1\pm1)}{4}\ketbra{g}{g} \int_{-\infty}^{\infty} \dd \tau \int_{-\infty}^\tau \dd \tau' \eta(\tau)\eta(\tau') e^{-\ii\Omega (\tau-\tau')}(\left( W(\sx_A,\sx_A') + W(\sx_B,\sx_B')\right)
    \\
    \tr_{\hat{\psi}}\left( \bra{\pm}\hat{U}^{(1)}\ket{\psi} \bra{\psi}\hat{U}^{(1)\dagger}\ket{\pm} \right) &= \frac{\lambda^2}{4}\ketbra{e}{e} \int_{-\infty}^{\infty} \dd \tau \int_{-\infty}^{\infty} \dd \tau' \eta(\tau)\eta(\tau') e^{\ii\Omega (\tau-\tau')} \nonumber \\
    &\quad \times \left( W(\sx_A',\sx_A) + W(\sx_B',\sx_B) \pm (W(\sx_B',\sx_A) + W(\sx_A',\sx_B)) \right)
\end{align}

To compute the transition probabilities, we must then add the relevant contributions together. In particular, the probability of finding the detector in the ground state corresponds to
\begin{align}
    P_G^{(\pm)} &= \tr_{g,\hat{\psi}}\left( \bra{\pm}\hat{U}^{(0)}\ket{\psi} \bra{\psi}\hat{U}^{(0)\dagger}\ket{\pm} + \bra{\pm}\hat{U}^{(0)}\ket{\psi} \bra{\psi}\hat{U}^{(2)\dagger}\ket{\pm} + \bra{\pm}\hat{U}^{(2)}\ket{\psi} \bra{\psi}\hat{U}^{(0)\dagger}\ket{\pm} \right)
    \nonumber \\
    &= \frac{1\pm1}{2} \left( 1-\frac{\lambda^2}{2} \int_{-\infty}^{\infty} \dd \tau \int_{-\infty}^\infty \dd \tau' \eta(\tau)\eta(\tau') e^{-\ii\Omega (\tau-\tau')}(\left( W(\sx_A,\sx_A') + W(\sx_B,\sx_B')\right) \right),
\end{align}
where we performed the substitutions $\tau\to-\tau$ and $\tau'\to-\tau'$ to rewrite the 0-2 term in such a way as to add it to the 2-0 term.

Performing a similar substitution in the last term of the 1-1 contribution and otherwise performing the substitution $\tau\leftrightarrow\tau'$, we can obtain the excitation probability to be
\begin{align}
    P_E^{(\pm)} &= \tr_{e,\hat{\psi}}\left( \bra{\pm}\hat{U}^{(1)}\ket{\psi} \bra{\psi}\hat{U}^{(1)\dagger}\ket{\pm} \right)
    \nonumber \\
    &= \frac{\lambda^2}{4} \int_{-\infty}^{\infty} \dd \tau \int_{-\infty}^{\infty} \dd \tau' \eta(\tau)\eta(\tau') e^{\ii\Omega (\tau-\tau')}
     \left( W(\sx_A,\sx_A') + W(\sx_B,\sx_B') \pm 2W(\sx_A,\sx_B')) \right)
\end{align}

Notice that we can then write both of these much more succinctly as
\begin{align}
    P_G^{(\pm)} &= \frac{1\pm1}{2} \left(1-\frac{\lambda^2}{2}(P_A+P_B)\right)
    \\
    P_E^{(\pm)} &= \frac{\lambda^2}{4}(P_A+P_B\pm2L_{AB})
\end{align}
where $P_D$ is the transition probability of an accelerated detector in a single copy of the cylindrically identified Minkowksi spacetime, while $L_{AB}$ is the contribution to the transition probability that arises from interference behaviours between the spacetimes. 

More explicitly, these are given in their simplest forms by
\begin{equation}
    P_D := \int_{-\infty}^{\infty} \dd \tau \int_{-\infty}^{\infty} \dd \tau' \eta(\tau)\eta(\tau') e^{\ii\Omega (\tau-\tau')} W(\sx_D,\sx_D'), 
\end{equation}
and
\begin{equation}
    L_{AB} := \int_{-\infty}^{\infty} \dd \tau \int_{-\infty}^{\infty} \dd \tau' \eta(\tau)\eta(\tau') e^{\ii\Omega (\tau-\tau')}
     W(\sx_A,\sx_B')
\end{equation}
which we express  in more detail below.

\subsubsection{Some comments on normalization}

The  final state of the detector after having the control state measured in the superposition basis is
\begin{equation}
    \rho^{(\pm)} = \begin{bmatrix}
        P_G^{(\pm)} & 0 \\
        0 & P_E^{(\pm)}
    \end{bmatrix}.
\end{equation}
This is not normalized, because we are restricting ourselves to the subset of outcomes \emph{conditioned} on the superposition measurement.
In order to normalize this state, we would have to consider something like
\begin{equation}
    \rho^{(\pm)}_{\rm norm} = \frac{1}{P_G^{(\pm)}+P_E^{(\pm)}} \begin{bmatrix}
        P_G^{(\pm)} & 0 \\
        0 & P_E^{(\pm)}
    \end{bmatrix}.
\end{equation}

If, on the other hand we had traced over the control system, rather than performing the measurement, we would be left with the state
\begin{equation}
    \rho = \begin{bmatrix}
        P_G^{(+)}+P_G^{(-)} & 0 \\
        0 & P_E^{(+)}+P_E^{(-)}
    \end{bmatrix} = \begin{bmatrix}
        1-\frac{\lambda^2}{2}(P_A+P_B) & 0 \\
        0 & \frac{\lambda^2}{2}(P_A+P_B)
    \end{bmatrix}
\end{equation}
which has no cross correlation terms. Instead, we are left with a (classical) mixture of the transition probability in each spacetime.

\section{Calculating $P_D, L_{AB}$}\label{appendix:p_d,l_ab}
The geodesic distance in Rindler coordinates from Eq. (\ref{eq:geodesic-distance-rindler}) can be computed by letting $r=r'=1/a$ as follows
\begin{equation}
    \begin{split}
        \sigma (\sx, \sx ') &= (t-t')^2 - (x-x')^2 - (y-y')^2 -(z-z')^2\\
        &= r^2(\sinh(a\tau)-\sinh(a\tau'))^2 -r^2(\cosh(a\tau)-\cosh(a\tau'))^2 -(y-y')^2 -(z-z')^2 \\
        &= r^2\left[4\cosh^2(\frac{a}{2}(\tau+\tau'))\sinh^2(\frac{a}{2}(\tau-\tau'))\right]-r^2\left[4\sinh^2(\frac{a}{2}(\tau+\tau'))\sinh^2(\frac{a}{2}(\tau-\tau'))\right]
         -(y-y')^2 -(z-z')^2 \\
        &= 4r^2\sinh^2(\frac{a}{2}(\tau-\tau'))\left[\cosh^2(\frac{a}{2}(\tau+\tau'))-\sinh^2(\frac{a}{2}(\tau+\tau'))\right] -(y-y')^2 -(z-z')^2\\
        &= \frac{4}{a^2}\sinh^2\left(\frac{a}{2}(\tau-\tau') \right) -(y-y')^2 -(z-z')^2.
    \end{split}
\label{app_eq:geodesic_explanation}
\end{equation}
The Wightman function  \eqref{eq:W_R} is 
straightforwardly computed using
\eqref{app_eq:geodesic_explanation}, with 
\begin{equation} 
        \text{sgn}(t-t') = \text{sgn}\left(\sinh(a\tau)- \sinh(a\tau')\right) = \text{sgn}\left[2\sinh(\frac{a}{2}(\tau-\tau'))\cosh(\frac{a}{2}(\tau+\tau'))\right]  = \text{sgn}\left(\tau-\tau'\right),
\label{app_eq:sgn_rindler}
\end{equation}
provided $a>0$. 

Next we write
\begin{equation}
    \begin{split}
        \delta (\sigma (J^n_{0_D} \sx, J^m_{0_D} \sx ')) &= \delta \left[ \frac{4}{a^2}\sinh^2\left(\frac{a}{2}(\tau-\tau')\right)-l_D^2(n-m)^2 \right] = \delta \left[ \frac{4}{a^2}\sinh^2\left(\frac{a}{2}s\right)-l_D^2(n-m)^2 \right],
    \end{split}
    \label{eq:delta_resolve}
\end{equation}
where  $s = \tau - \tau'$. The argument $f(s) = \frac{4}{a^2}\sinh^2\left(\frac{a}{2}s\right)-l_D^2(n-m)^2$ of the delta function vanishes for 
\begin{equation}
s_\pm = \frac{2}{a}\sinh^{-1}\left[\pm\frac{a}{2}l_D(n-m)\right]  = \pm\frac{2}{a}\sinh^{-1}\left[\frac{a}{2}l_D(n-m)\right],
\end{equation}
and its deriviative is
\begin{equation}
    \begin{split}
        &\frac{d}{ds}f(s)\vert_{s_\pm} = \frac{4}{a}\sinh\left(\frac{a}{2}s_\pm\right)\cosh\left(\frac{a}{2}s_\pm\right)
        = \pm2\,l_D(n-m)\sqrt{1+\frac{a^2}{4}l_D^2(n-m)^2},
    \end{split}
\end{equation}
thereby yielding
\begin{equation}
    \delta (\sigma (J^n_{0_D} \sx, J^m_{0_D} \sx ')) = \frac{\delta\left(s - \frac2a\sinh^{-1}(\frac{a}2l_D(n-m))\right)+\delta\left(s + \frac2a\sinh^{-1}(\frac{a}2l_D(n-m))\right)}{2\,l_D|n-m|\sqrt{1+\frac{a^2}{4}l_D^2(n-m)^2}} \text{ for } n\neq m.
\end{equation}

To compute $L_{AB}$ we need  $\delta (\sigma (J^n_{0_A} \sx, J^m_{0_B} \sx '))$. A similar calculation gives
\begin{equation}
    \delta (\sigma (J^n_{0_A} \sx, J^m_{0_B} \sx ')) = \frac{\delta\left(s - \frac2a\sinh^{-1}(\frac{a}2(l_A n-l_B m))\right)+\delta\left(s + \frac2a\sinh^{-1}(\frac{a}2(l_A n-l_B m))\right)}{2\,|l_A n-l_B m|\sqrt{1+\frac{a^2}{4}(l_A n-l_B m)^2}} \text{ for } l_A n\neq l_B m.
\end{equation}

To compute $P_D$ in Eq. \eqref{eq:transition probability single detector}, we take $\eta = 1$ and insert   \eqref{eq:Wightman_D} as follows:
\begin{equation}
    \begin{split}
        W_{R_0}^{(D)} (\sx, \sx ') &= \frac{1}{\mathcal{N}} \sum_{n,m} \eta^n \eta^m \left[ \frac{\text{sgn}(\tau-\tau') \delta (\sigma _\mathcal{R}(J^n_{0_D} \sx, J^m_{0_D} \sx '))}{4\pi i}- \frac{1}{4 \pi ^2 \sigma _\mathcal{R} (J^n_{0_D} \sx, J^m_{0_D} \sx ')}\right] \\
        &= \underbrace{\frac{1}{\mathcal{N}} \sum_{n=m}\left[ \frac{\text{sgn}(s)\delta \left[ \frac{4}{a^2}\sinh^2\left(\frac{a}{2}s\right)\right]}{4\pi i} - \frac{1}{4\pi^2 [\frac{4}{a^2}\sinh^2\left(\frac{a}{2}s\right)]}\right]}_{W_\mathcal{R}(s)} \\
        &\,\,\,+ \frac{1}{\mathcal{N}} \sum_{n\neq m}\left[ \frac{\text{sgn}(s) \delta \left[ \frac{4}{a^2}\sinh^2\left(\frac{a}{2}s\right)-l_D^2(n-m)^2 \right]}{4\pi i} - \frac{1}{4\pi^2 \left[ \frac{4}{a^2}\sinh^2\left(\frac{a}{2}s\right)-l_D^2(n-m)^2 \right]} \right] \\
        &= W_\mathcal{R}(s) + \frac{1}{\mathcal{N}} \sum_{n\neq m}\left[ \frac{\text{sgn}(s) \delta \left[ \frac{4}{a^2}\sinh^2\left(\frac{a}{2}s\right)-l_D^2(n-m)^2 \right]}{4\pi i} - \frac{1}{4\pi^2 \left[ \frac{4}{a^2}\sinh^2\left(\frac{a}{2}s\right)-l_D^2(n-m)^2 \right]} \right]\\
        \implies W_{R_0}^{(D)} (\sx, \sx ') &= W_{R_0}^{(D)} (s)
    \end{split}
    \label{}
\end{equation}
where again $s = \tau - \tau'$. The transition probability is given by
\begin{equation}
    \begin{split}
        P_D &= \int_{-\infty}^\infty d\tau \int_{-\infty}^\infty d\tau' \, e^{-\frac{\tau^2}{2\sigma^2}}e^{-\frac{\tau'^2}{2\sigma^2}} e^{-i\Omega(\tau-\tau')} \, W_{R_0}^{(D)} (\sx, \sx '),\\
        &= \int_{-\infty}^\infty d u \int_{-\infty}^\infty d s \, e^{-\frac{u^2}{2\sigma^2}}e^{-\frac{(u-s)^2}{2\sigma^2}} e^{-i\Omega s} \, W_{R_0}^{(D)} (s),\\
        &= \sqrt{\pi} \sigma \int_{-\infty}^\infty d s \, e^{-\frac{s^2}{4\sigma^2}} e^{-i\Omega s} \, W_{R_0}^{(D)}(s),
    \end{split}
    \label{}
\end{equation}
having performed the $du$ integral in the last line. Expanding the Wightman function, we get
\begin{equation}
    \begin{split}
        P_D &= \sqrt{\pi} \sigma \int_{-\infty}^\infty d s \, e^{-\frac{s^2}{4\sigma^2}} e^{-i\Omega s} \, \left[ W_\mathcal{R}(s) + \frac{1}{\mathcal{N}} \sum_{n\neq m}\left[ \frac{\text{sgn}(s) \delta \left[ \frac{4}{a^2}\sinh^2\left(\frac{a}{2}s\right)-l_D^2(n-m)^2 \right]}{4\pi i} \right.\right.\\
        & \hspace{130pt} - \left.\left.  \frac{1}{4\pi^2 \left[ \frac{4}{a^2}\sinh^2\left(\frac{a}{2}s\right)-l_D^2(n-m)^2 \right]} \right] \right],\\ \\
        &= P_\mathcal{R} + \frac{\sqrt{\pi} \sigma}{\mathcal{N}} \sum_{n\neq m} \int_{-\infty}^\infty d s \, e^{-\frac{s^2}{4\sigma^2}} e^{-i\Omega s} \left[ \underbrace{\frac{\text{sgn}(s) \delta \left[ \frac{4}{a^2}\sinh^2\left(\frac{a}{2}s\right)-l_D^2(n-m)^2 \right]}{4\pi i}}_{I_1} \right.\\
        & \hspace{180pt} - \left. \underbrace{\frac{1}{4\pi^2 \left[ \frac{4}{a^2}\sinh^2\left(\frac{a}{2}s\right)-l_D^2(n-m)^2 \right]}}_{I_2} \right].
    \end{split}
    \label{}
\end{equation}
The first term can be written as
\begin{equation}
    \begin{split}
        P_\mathcal{R} &= \sqrt{\pi} \sigma \int_{-\infty}^\infty d s \, e^{-\frac{s^2}{4\sigma^2}} e^{-i\Omega s} \, W_\mathcal{R}(s) \\
        &= \frac{\sqrt{\pi} \sigma}{\mathcal{N}}\sum_{n=m} \int_{-\infty}^\infty d s \, e^{-\frac{s^2}{4\sigma^2}} e^{-i\Omega s} \, \left[ \frac{\text{sgn}(s)\delta \left[ \frac{4}{a^2}\sinh^2\left(\frac{a}{2}s\right)\right]}{4\pi i} - \frac{1}{4\pi^2 [\frac{4}{a^2}\sinh^2\left(\frac{a}{2}s\right)]}\right]\\
        &= \frac{\sqrt{\pi}\sigma}{\mathcal{N}}\frac{1}{4\pi} \sum_{n=m}\left[ T_1 + T_2 \right],
    \end{split}
    \label{app_eq:P_R}
\end{equation}
where  
\begin{equation}
    \begin{split}
        T_1 &:= -i \int_{-\infty}^\infty d s \, e^{-\frac{s^2}{4\sigma^2}} e^{-i\Omega s}\, \text{sgn}(s)\delta \left( \frac{4}{a^2}\sinh^2\left(\frac{a}{2}s\right)\right),\\
        T_2 &:= -\frac{1}{\pi}\int_{-\infty}^\infty d s \, e^{-\frac{s^2}{4\sigma^2}} e^{-i\Omega s}\, \frac{a^2}{ 4\sinh^2\left(\frac{a}{2}s\right)}.
    \end{split}
\end{equation}
Let us examine $T_1$ first. Observe that for a well-behaved test function $f(x)$, we have 
\begin{equation}
    \begin{split}
        \text{PV} \int_{-\infty}^\infty dx\,f(x)\, \text{sgn}(x)\, \delta \left( \frac{4}{a^2}\sinh^2\left(\frac{a}{2}x\right)\right) &= \lim_{r\to 0}\text{PV} \int_{-\infty}^\infty dx\, f(x)\, \text{sgn}(x)\, \delta \left( \frac{4}{a^2}\sinh^2\left(\frac{a}{2}x\right) - r^2\right)\\
        &= \lim_{r\to 0}\text{PV} \int_{-\infty}^\infty dx\, f(x)\, \text{sgn}(x)\, \frac{\delta \left( \frac{2}{a}\sinh \left(\frac{a}{2}x\right) - r\right) + \delta \left( \frac{2}{a}\sinh \left(\frac{a}{2}x\right) + r\right)}{2|r|}\\
        &= \lim_{r\to 0}\text{PV} \int_{-\infty}^\infty dx\, f(x)\, \text{sgn}(x)\,\frac{\delta \left( x - \frac{2}{a}\sinh^{-1}\left(\frac{a}{2}r\right)\right) + \delta \left( x + \frac{2}{a}\sinh^{-1}\left(\frac{a}{2}r\right)\right)}{2|r| \sqrt{a^2 r^2 /4 + 1}}\\
        &= \lim_{r\to 0} \frac{f\left(\frac{2}{a}\sinh^{-1}\left(\frac{a}{2}r\right)\right) - f\left(-\frac{2}{a}\sinh^{-1}\left(\frac{a}{2}r\right)\right)}{2|r| \sqrt{a^2 r^2 /4 + 1}}\\
        &= \lim_{r\to 0} \frac{\frac{df}{dx}(\frac2a \sinh^{-1}(\frac{a}{2}r)).\frac{2}{a}.\frac{1}{\sqrt{a^2r^2/4+1}}.\frac{a}{2} - \frac{df}{dx}(-\frac2a \sinh^{-1}(\frac{a}{2}r)).\frac{-2}{a}.\frac{1}{\sqrt{a^2r^2/4+1}}.\frac{a}{2}}{2\sqrt{a^2 r^2 /4 + 1} + 2|r|\frac{1}{2}(a^2r^2/4+1)^{-1/2} a^2 r/2}\\
        &= \frac{d}{dx}f(x)\Bigr|_{x=0} =: f'(0).
    \end{split}
\end{equation}
Thus, taking $f(s) = e^{-\frac{s^2}{4\sigma^2}} e^{-i\Omega s}$, $T_1$ simplifies to
\begin{equation}
        T_1 = -i \frac{d}{ds} e^{-\frac{s^2}{4\sigma^2}} e^{-i\Omega s} \Bigr|_{s=0} = -i \left(-\frac{2s}{4\sigma^2}-i\Omega\right) e^{-\frac{s^2}{4\sigma^2}} e^{-i\Omega s} \Bigr|_{s=0} = i^2\Omega = -\Omega.
\end{equation}
To evaluate $T_2$, we make use of the identity
\begin{align}
    % \begin{split}
        \text{PV} \int_{-\infty}^\infty dx\, \frac{f(x)}{x^2} =  \int_{0}^\infty dx \, \frac{f(x)+f(-x)-2f(0)}{x^2}.
    % \end{split}
\end{align}
Thus, taking $f(s) = e^{-\frac{s^2}{4\sigma^2}} e^{-i\Omega s} \frac{s^2}{\sinh^2(\frac{a}{2}s)}$, $T_2$ simplifies to
\begin{align}
   % \begin{split}
        T_2 &= -\frac{a^2}{4\pi} \int_{0}^{\infty} ds\, \frac{e^{-\frac{s^2}{4\sigma^2}} e^{-i\Omega s} \frac{s^2}{\sinh^2(\frac{a}{2}s)} + e^{-\frac{s^2}{4\sigma^2}} e^{i\Omega s} \frac{s^2}{\sinh^2(\frac{a}{2}s)} - 2\left(\lim_{s\to 0} \frac{s}{\sinh(\frac{a}{2}s)}\right)^2}{s^2}\\
        &= -\frac{a^2}{4\pi} \int_{0}^{\infty} ds\, \frac{ \frac{s^2}{\sinh^2(\frac{a}{2}s)} e^{-\frac{s^2}{4\sigma^2}} \left[ e^{-i\Omega s}  + e^{i\Omega s} \right]  - 2\left(\frac{2}{a}\right)^2}{s^2}\\
        &= -\frac{a^2}{2\pi} \int_{0}^{\infty} ds\,\left[  \frac{e^{-\frac{s^2}{4\sigma^2}} \cos(\Omega s)}{\sinh^2(\frac{a}{2}s)}  - \frac{4}{a^2 s^2}\right]
    %\end{split}
\end{align}
Hence  $P_\mathcal{R}$ in \eqref{app_eq:P_R} is given by
\begin{equation}
    \begin{split}
        P_\mathcal{R} &= \frac{\sqrt{\pi}\sigma}{\mathcal{N}}\frac{1}{4\pi} \sum_{n=m}\left[ T_1 + T_2 \right] = \frac{\sqrt{\pi}\sigma}{\mathcal{N}}\frac{1}{4\pi} \sum_{n=m} \left[ -\Omega -\frac{a^2}{2\pi} \int_{0}^{\infty} ds\,\left(  \frac{e^{-\frac{s^2}{4\sigma^2}} \cos(\Omega s)}{\sinh^2(\frac{a}{2}s)}  - \frac{4}{a^2 s^2}\right) \right]\\
        &= \frac{\sqrt{\pi}\sigma}{\mathcal{N}} \sum_{n=m} \left[ -\frac{\Omega}{4\pi} -\frac{a^2}{8\pi^2} \int_{0}^{\infty} ds\,\left(  \frac{e^{-\frac{s^2}{4\sigma^2}} \cos(\Omega s)}{\sinh^2(\frac{a}{2}s)}  - \frac{4}{a^2 s^2}\right) \right].
    \end{split}
\end{equation}

For the second term $I_1$ we get
\begin{equation}
    \begin{split}
        I_1 &= \sum_{n\neq m} \int_{-\infty}^\infty d s \, e^{-\frac{s^2}{4\sigma^2}} e^{-i\Omega s} \left[ \frac{\text{sgn}(s) \delta \left[ \frac{4}{a^2}\sinh^2\left(\frac{a}{2}s\right)-l_D^2(n-m)^2 \right]}{4\pi i}\right]\\
        &= \frac{1}{4\pi i} \sum_{n\neq m} \int_{-\infty}^\infty d s \, e^{-\frac{s^2}{4\sigma^2}} e^{-i\Omega s} \text{sgn}(s) \delta \left[ \frac{4}{a^2}\sinh^2\left(\frac{a}{2}s\right)-l_D^2(n-m)^2 \right]\\
        &= \frac{1}{4\pi i} \sum_{n\neq m} \int_{-\infty}^\infty d s \, e^{-\frac{s^2}{4\sigma^2}} e^{-i\Omega s} \text{sgn}(s)\, \frac{\delta\left(s - \frac2a\sinh^{-1}(\frac{a}2l_D(n-m))\right)+\delta\left(s + \frac2a\sinh^{-1}(\frac{a}2l_D(n-m))\right)}{2\,l_D|n-m|\sqrt{1+\frac{a^2}{4}l_D^2(n-m)^2}}\\
        &= \frac{1}{4\pi i} \sum_{n\neq m} \int_{-\infty}^\infty d s \, e^{-\frac{s^2}{4\sigma^2}} e^{-i\Omega s} \text{sgn}(s)\, \frac{\delta\left(s - \frac2a\sinh^{-1}(\frac{a}2 k_D)\right)+\delta\left(s + \frac2a\sinh^{-1}(\frac{a}2 k_D)\right)}{2\,l_D|n-m|\sqrt{1+\frac{a^2}{4}k_D^2}}\\
        &= \frac{1}{4\pi i} \sum_{n\neq m} e^{-\frac{(\sinh^{-1}(\frac a2 k_D))^2}{a^2\sigma^2}} \left[ \frac{e^{-i\Omega \frac2a\sinh^{-1}(\frac{a}2 k_D)} \text{sgn}(\frac2a\sinh^{-1}(\frac{a}2 k_D)) + e^{i\Omega \frac2a\sinh^{-1}(\frac{a}2 k_D)} \text{sgn}(-\frac2a\sinh^{-1}(\frac{a}2 k_D))}{2\,l_D|n-m|\sqrt{1+\frac{a^2}{4}k_D^2}}\right]\\   
        &= \frac{1}{4\pi i} \sum_{n>m} e^{-\frac{(\sinh^{-1}(\frac a2 k_D))^2}{a^2\sigma^2}} \left( \frac{e^{-i\Omega \frac2a\sinh^{-1}(\frac{a}2 k_D)} - e^{i\Omega \frac2a\sinh^{-1}(\frac{a}2 k_D)}}{2\,l_D(n-m)\sqrt{1+\frac{a^2}{4}k_D^2}}\right) \\
        &\,\,\,\,\,\,+ \frac{1}{4\pi i} \sum_{n<m} e^{-\frac{(\sinh^{-1}(\frac a2 k_D))^2}{a^2\sigma^2}} \left( \frac{- e^{-i\Omega \frac2a\sinh^{-1}(\frac{a}2 k_D)} + e^{i\Omega \frac2a\sinh^{-1}(\frac{a}2 k_D)}}{-2\,l_D(n-m)\sqrt{1+\frac{a^2}{4}k_D^2}}\right)\\
        &= \frac{1}{4\pi} \sum_{n>m} e^{-\frac{(\sinh^{-1}(\frac a2 k_D))^2}{a^2\sigma^2}} \left( \frac{- \sin\left(\frac{2\Omega}{a}\sinh^{-1}(\frac{a}{2}l_D(n-m))\right)}{l_D(n-m)\sqrt{1+\frac{a^2}{4}k_D^2}}\right)\\
        &\,\,\,\,\,\,+ \frac{1}{4\pi } \sum_{n<m} e^{-\frac{(\sinh^{-1}(\frac a2 k_D))^2}{a^2\sigma^2}} \left( \frac{ \sin\left(\frac{2\Omega}{a}\sinh^{-1}(\frac{a}{2}l_D(n-m))\right)}{-l_D(n-m)\sqrt{1+\frac{a^2}{4}k_D^2}}\right)\\
        &= \frac{1}{4\pi} \sum_{n>m}e^{-\frac{(\sinh^{-1}(\frac a2 k_D))^2}{a^2\sigma^2}} \left[
        \frac{- \sin\left(\frac{2\Omega}{a}\sinh^{-1}(\frac{a}{2}l_D(n-m))\right)}{l_D(n-m)\sqrt{1+\frac{a^2}{4}k_D^2}}
        + \frac{\sin\left(\frac{2\Omega}{a}\sinh^{-1}(\frac{a}{2}l_D(n-m))\right)}{-l_D(n-m)\sqrt{1+\frac{a^2}{4}k_D^2}}
        \right]\\
        &= -\frac{1}{2\pi} \sum_{n>m}e^{-\frac{(\sinh^{-1}(\frac a2 k_D))^2}{a^2\sigma^2}} \frac{ \sin\left(\frac{2\Omega}{a}\sinh^{-1}(\frac{a}{2}l_D(n-m))\right)}{l_D(n-m)\sqrt{1+\frac{a^2}{4}k_D^2}}\\
        &= - \sum_{n>m}\frac{1}{2\pi k_D} \frac{e^{-\frac{(\sinh^{-1}(a k_D/2))^2}{a^2\sigma^2}}}{\sqrt{\frac{a^2}{4}k_D^2+1}}\sin\left(\frac{2\Omega}{a}\sinh^{-1}\left(\frac{a}{2}k_D\right)\right)
    \end{split}
\end{equation}
where $k_D = l_D(n-m)$.

For $I_2$
we find
\begin{equation}
    \begin{split}
        I_2 &= -\frac{1}{4\pi^2}
        \sum_{n\neq m} \int_{-\infty}^\infty d s \,\left[ \frac{e^{-\frac{s^2}{4\sigma^2}} e^{-i\Omega s}}{\frac{4}{a^2}\sinh^2\left(\frac{a}{2}s\right)-l_D^2(n-m)^2}\right]\\
        &= -\frac{1}{4\pi^2}
        \sum_{n\neq m} \int_{-\infty}^\infty d s \int_{-\infty}^{\infty} d s' \, \delta (s-s') \left[ \frac{e^{-\frac{s'^2}{4\sigma^2}} e^{-i\Omega s'}}{\frac{4}{a^2}\sinh^2\left(\frac{a}{2}s\right)-k_D^2}\right]\\
        &= -\frac{1}{4\pi^2}
        \sum_{n\neq m} \int_{-\infty}^\infty d s \int_{-\infty}^{\infty} d s' \, \left(\frac{1}{2\pi}\int_{-\infty}^{\infty} dz e^{-i z (s-s')}\right) \left[ \frac{e^{-\frac{s'^2}{4\sigma^2}} e^{-i\Omega s'}}{\frac{4}{a^2}\sinh^2\left(\frac{a}{2}s\right)-k_D^2}\right]\\
        &= -\frac{1}{8\pi^3} \sum_{n\neq m}  \int_{-\infty}^\infty d z \left(\int_{-\infty}^\infty d s' e^{-i(\Omega - z)s'}e^{-\frac{s'^2}{4\sigma^2}}\right)
        \left(\int_{-\infty}^{\infty} d s \frac{e^{-i z s}}{\frac{4}{a^2}\sinh^2\left(\frac{a}{2}s\right)-k_D^2}\right)\\
        &= -\frac{1}{8\pi^3} \sum_{n\neq m}  \int_{-\infty}^\infty d z \left(2\sqrt{\pi}\sigma e^{-(\Omega-z)^2\sigma^2}\right) \cdot I_3
    \end{split}
\end{equation}
where $I_3$ can be computed as follows
\begin{equation}
    \begin{split}
        I_3 &= \int_{-\infty}^{\infty} d s\, \frac{e^{-i z s}}{\frac{4}{a^2}\sinh^2\left(\frac{a}{2}s\right)-k_D^2}\\
        &= \frac{a^2}{4}\int_{-\infty}^{\infty} d s\, \frac{e^{-i z s}}{\sinh^2\left(\frac{a}{2}s\right)-\frac{a^2}{4}k_D^2}\\
        &= \frac{a^2}{4}\int_{-\infty}^{\infty} d s\, \frac{e^{-i z s}}{\sinh^2\left(\frac{a}{2}s\right)-\sinh^2(t)}, \,\,\text{for}\,\, \sinh(t) = \frac{a}{2}k_D\\
        &=\frac{a^2}{4}\int_{-\infty}^{\infty} d s\, \frac{e^{-i z s}}{\left[\sinh\left(\frac{a}{2}s\right)-\sinh(t)\right]\left[\sinh\left(\frac{a}{2}s\right)+\sinh(t)\right]}\\
        &=\frac{a^2}{4}\frac{1}{2 \sinh(t)} \left[\int_{-\infty}^\infty d s\, \frac{e^{-izs}}{\sinh\left(\frac{a}{2}s\right)+\sinh(-t)} - \int_{-\infty}^\infty d s\, \frac{e^{-izs}}{\sinh\left(\frac{a}{2}s\right)+\sinh(t)}\right]\\
        &=\frac{a}{2}\frac{1}{2 \sinh(t)} \left[\int_{-\infty}^\infty d u\, \frac{e^{-iz\frac2au}}{\sinh\left(u\right)+\sinh(-t)} - \int_{-\infty}^\infty d u\, \frac{e^{-iz\frac2au}}{\sinh\left(u\right)+\sinh(t)}\right], \,\,\text{for}\,\, u = \frac{a}{2}s\\
        &= \frac{a}{2}\frac{1}{2 \sinh(t)} \left[
        -\frac{i\pi e^{i(-t)\frac2az}}{\sinh(\pi \frac2az)\cosh(-t)}\left(\cosh(\pi \frac2az) - e^{2i t\frac2az}\right)
        +\frac{i\pi e^{i t\frac2az}}{\sinh(\pi \frac2az)\cosh(t)}\left(\cosh(\pi \frac2az) - e^{-2i t\frac2az}\right)
        \right], \text{for}\,\, t>0\\
        &= \frac{a}{2}\frac{i\pi}{2 \sinh(t)\cosh(t)} 
        \left[\left(-e^{-i\frac2atz}+e^{i\frac2atz}\right) \frac{\cosh(\frac{2\pi}{a}z)}{\sinh(\frac{2\pi}{a}z)} + \frac{\left(e^{i\frac2atz}-e^{-i\frac2atz}\right)}{\sinh(\frac{2\pi}{a}z)}
        \right]\\
        &= \frac{a}{2}\frac{i\pi}{2 \sinh(t)\cosh(t)} 
        \left[2 i \sin(\frac2atz)\left(\frac{\cosh(\frac{2\pi}{a}z)+1}{\sinh(\frac{2\pi}{a}z)}\right)
        \right]\\
        &= \frac{a}{2}\frac{-\pi}{ \sinh(t)\cosh(t)} \left[\sin(\frac2atz)\left(\frac{2\cosh^2(\frac{\pi z}{a})}{2\sinh(\frac{\pi z}{a})\cosh(\frac{\pi z}{a})}\right)
        \right]\\
        &= \frac{a}{2}\frac{-\pi}{ \sinh(t)\cosh(t)} \left[\sin(\frac2atz)\coth(\frac{\pi z}{a})
        \right]\\
        &= \frac{-\pi}{k_D \sqrt{\frac{a^2}{4}k_D^2 + 1}} \coth(\frac{\pi z}{a})
        \sin(\frac{2z}{a}\sinh^{-1}\left(\frac{a}{2}k_D\right)).
    \end{split}
\end{equation}
Thus, $I_2$ is given by
\begin{equation}
    \begin{split}
        I_2 &= -\frac{1}{8\pi^3} \sum_{n\neq m}  \int_{-\infty}^\infty d z \left(2\sqrt{\pi}\sigma e^{-(\Omega-z)^2\sigma^2}\right) \cdot \frac{-\pi}{k_D \sqrt{\frac{a^2}{4}k_D^2 + 1}} \coth(\frac{\pi z}{a})
        \sin(\frac{2z}{a}\sinh^{-1}\left(\frac{a}{2}k_D\right))\\
        &= \sum_{n\neq m} \frac{\sigma}{4\pi\sqrt{\pi} k_D \sqrt{\frac{a^2}{4}k_D^2 + 1}} \int_{-\infty}^\infty dz\, e^{-(\Omega-z)^2\sigma^2} \coth(\frac{\pi z}{a})
        \sin(\frac{2z}{a}\sinh^{-1}\left(\frac{a}{2}k_D\right)).
    \end{split}
\end{equation}
Finally combining $P_\mathcal{R}, I_1$ and $I_2$, we get the final form of $P_D$ as shown below. Similar calculation can be done for $L_{AB}$ to give the following formula 
\cite{FAZM_2023_flat, ARHSmith2017}
\begin{align}
    \begin{split}
        P_D = \frac{\sqrt{\pi}\sigma}{\mathcal{N}} &\left[\, \, \,\sum_{n=m}\left( -\frac{\Omega}{4\pi} - \frac{a^2}{8\pi^2} \int_0^\infty ds \left( \frac{e^{-\frac{s^2}{4\sigma^2}} \cos(\Omega s)}{\sinh^2(\frac{as}{2})}-\frac{4}{a^2s^2}\right) \right)\right.\\
        & - \left.\sum_{n>m} \frac{1}{2\pi k_D} \frac{e^\frac{- (\sinh^{-1}(ak_D/2))^2}{a^2\sigma^2}}{\sqrt{\frac{a^2}{4}k_D^2 + 1}} \sin(\frac{2\Omega}{a} \sinh^{-1}(\frac{a}{2}k_D))\right.\\
        &+ \left. \sum_{n\neq m} \frac{\sigma}{4\pi\sqrt{\pi}k_D} \int_{-\infty}^{\infty} dz\, \, \frac{e^{-(\Omega-z)^2\sigma^2}\coth(\frac{\pi z}{a})}{\sqrt{\frac{a^2}{4}k_D^2+1}} \sin(\frac{2z}{a} \sinh^{-1}(\frac{a}{2}k_D))\right] 
        \label{app_eq:P_D form} 
    \end{split}
\\
    \begin{split}
        L_{AB} = \frac{\sqrt{\pi}\sigma}{\mathcal{N}}&\left[\, \, \,\sum_{l_A n=l_B m}\left( -\frac{\Omega}{4\pi} - \frac{a^2}{8\pi^2} \int_0^\infty ds \left( \frac{e^{-\frac{s^2}{4\sigma^2}} \cos(\Omega s)}{\sinh^2(\frac{as}{2})}-\frac{4}{a^2s^2}\right) \right)\right.\\
        & - \left.\sum_{l_A n>l_B m} \frac{1}{2\pi k_{AB}} \frac{e^\frac{- (\sinh^{-1}(ak_{AB}/2))^2}{a^2\sigma^2}}{\sqrt{\frac{a^2}{4}k_{AB}^2 + 1}} \sin(\frac{2\Omega}{a} \sinh^{-1}(\frac{a}{2}k_{AB}))\right.\\
        &+ \left. \sum_{l_A n\neq l_B m} \frac{\sigma}{4\pi\sqrt{\pi} k_{AB}} \int_{-\infty}^{\infty} dz\, \, \frac{e^{-(\Omega-z)^2\sigma^2}\coth(\frac{\pi z}{a})}{\sqrt{\frac{a^2}{4}k_{AB}^2+1}} \sin(\frac{2z}{a} \sinh^{-1}(\frac{a}{2}k_{AB}))
        \right] \label{app_eq:L_AB form} 
    \end{split}
\end{align}
where $k_D = l_D(n-m)$, $k_{AB} = (l_A n-l_B m)$, $\mathcal{N} = \sum_n \eta^{2n}$ and $s = \tau - \tau'$ (the proper time difference).

\section{Comparison with Static Detector}\label{appendix:compare with a=0}

Taking the limit $a\to 0$
in \eqref{app_eq:P_D form} 
and \eqref{app_eq:L_AB form} 
we find
\begin{align}
    \begin{split}
        \lim_{a\to 0} P_D = \frac{\sqrt{\pi}\sigma}{\mathcal{N}}&\left[\, \, \,\sum_{n=m}\left( -\frac{\Omega}{4\pi} - \frac{1}{2\pi^2} \int_0^\infty ds \left( \frac{ e^{-\frac{s^2}{4\sigma^2}} \cos(\Omega s)}{s^2}-\frac{1}{s^2}\right) \right)\right.\\
        & - \left.\sum_{n>m} \frac{e^\frac{- l_D^2(n-m)^2}{4\sigma^2}}{2\pi l_D (n-m)} \sin(\Omega l_D(n-m))\right.\\
        &+ \left. \sum_{n\neq m} \frac{\sigma}{4\pi\sqrt{\pi}l_D (n-m)}\int_{-\infty}^{\infty} dz\, \, e^{-(\Omega-z)^2\sigma^2}\text{sgn}(z) \sin(z l_D (n-m))\right] 
        \label{app_eq:P_D formB} 
    \end{split}
\\
    \begin{split}
        \lim_{a\to 0} L_{AB} = \frac{\sqrt{\pi}\sigma}{\mathcal{N}}&\left[\, \, \,\sum_{l_A n=l_B m}\left( -\frac{\Omega}{4\pi} - \frac{1}{2\pi^2} \int_0^\infty ds \left( \frac{ e^{-\frac{s^2}{4\sigma^2}} \cos(\Omega s)}{s^2}-\frac{1}{s^2}\right) \right)\right.\\
        & - \left.\sum_{l_A n>l_B m}\frac{e^\frac{- (l_A n-l_B m)^2}{4\sigma^2}}{2\pi (l_A n-l_B m)} \sin(\Omega (l_A n-l_B m))\right.\\
        &+ \left. \sum_{l_A n\neq l_B m} \frac{\sigma}{4\pi\sqrt{\pi} (l_A n-l_B m)}\int_{-\infty}^{\infty} dz\, \, e^{-(\Omega-z)^2\sigma^2}\text{sgn}(z) \sin(z (l_A n-l_B m))
        \right] \label{app_eq:L_AB formB} 
    \end{split}
\end{align} which match  the
corresponding formulae for a static detector
in superposed cylindrically identified Minkowski spacetime 
\cite{FAZM_2023_flat}.
We plot 
$P_A, P_B, P_E^{(\pm)}$ and
$L_{AB}$ from
\eqref{app_eq:P_D formB}  and
\eqref{app_eq:L_AB formB} 
as functions of the ratio of the   characteristic circumferences $l_A \text{ and } l_B$ in
\autoref{fig:comparison-pe-length-ratio}, and of
the energy gap $\Omega\sigma$ in
\autoref{fig:pe-energy-gap:a}.  These   match  the plots 
previously obtained \cite{FAZM_2023_flat} for the static case.

\begin{figure}[H]
\centering
  \includegraphics[width=0.5\linewidth]{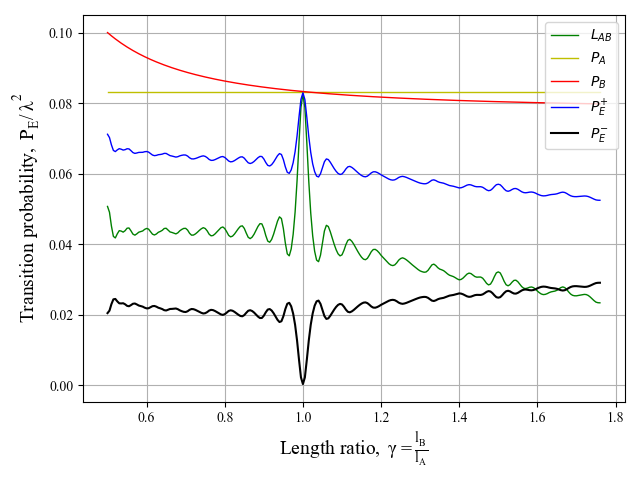}
  \captionof{figure}{Plots of $P_A$ (yellow), $P_B$ (red), $P_E^{(+)}$  (blue), $P_E^{(-)}$  (black),  and
$L_{AB}$ (green) as a function of the ratio of the two characteristic circumferences $l_A \text{ and } l_B$  for static detector with parameters $\sigma\Omega=0.01$, $l_A/\sigma = 10$ and varying $l_B/\sigma$ from $5$ to $17.5$.}
  \label{fig:comparison-pe-length-ratio}
% \end{minipage}%
% \hfill
% \begin{minipage}{.475\textwidth}
%   \centering
%   \includegraphics[width=1\linewidth]{images/final_n_40_a_0_dot.png}
%   \captionof{figure}{Plots of $P_A$ (yellow), $P_B$ (red), $P_E^{(+)}$  (blue), $P_E^{(-)}$  (black),  and
% $L_{AB}$ (green)  as a function of energy gap $\Omega\sigma$ for static detector with parameters $l_A/\sigma = 0.75, l_B/\sigma=0.25$.  }
%   \label{fig:comparison-pe-gamma-different-acc}
% \end{minipage}
\end{figure}

% \begin{figure}[t]
%     \centering
%     \includegraphics[width=0.45\linewidth]{images/final_n_20_a_0.0_dot_step_0.001.png}
%     \caption{Caption}
%     \label{fig:comparison fig:pe-length-ratio}
% \end{figure}

% \begin{figure}[t]
%     \centering
%     \includegraphics[width=0.45\linewidth]{images/final_n_40_a_0.0_dot.png}
%     \caption{Caption}
%     \label{fig:comparison fig:pe-length-ratio}
% \end{figure}

\section{Convergence of Image Sums in Transition Probability}\label{appendix:convergence of image sum}

To demonstrate numerical convergence of the image sums,
we compute the individual contributions  
  $P_E^{(+)}(n,m)$  
from \eqref{eq:P_D form}, \eqref{eq:L_AB form}   using Python \cite{Goel_Accelerated_Detector_in}. We pick $n,m$ up to some integer value $n_{\text{max}}$ and plot 
the resulting $P_E^{(+)}
(n_{\text{max}},n_{\text{max}})$
for different $n_{\text{max}}$ 
as a function of the ratio 
$l_A/\sigma/l_B/\sigma$ for
various accelerations and gaps. We find good convergence for 50 terms; including more terms introduces no numerically discernable changes.

\begin{figure}[H]
    \centering
    \captionsetup{justification=centering}
    \includegraphics[width=0.75\linewidth]{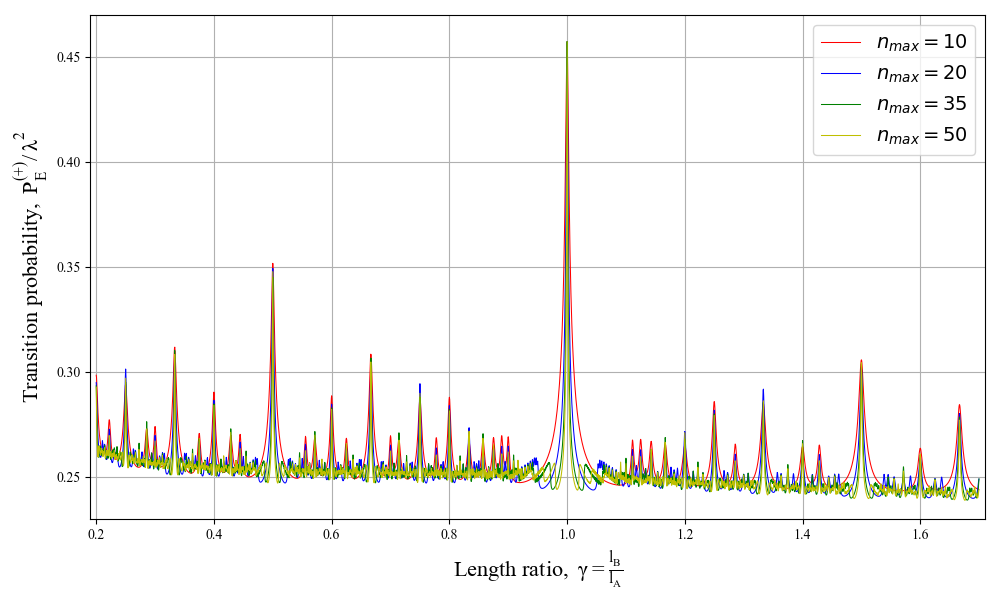}
    \caption{\justifying Plot of transition probability $P_E^{(+)}$ as a function of $\gamma$ for acceleration $a\sigma = 10$ with parameters $\Omega\sigma=0.01$, fixed $l_A/\sigma = 10$ and varying $l_B/\sigma$ from $2$ to $17$ with varied $n_\text{max}$.}
    \label{fig:convergence_image_sum}
\end{figure}
\autoref{fig:convergence_image_sum} provides a typical example. We can see that the transition probability converges to some value as $n_\text{max}$ increases. While the prominent resonance peaks converge, the number of peaks of lesser amplitudes increase with increasing $n_\text{max}$.
\end{widetext}
\bibliography{main}
\end{document}